\documentclass[preprint,prd,aps,showpacs,showkeys,nofootinbib]{revtex4}
\usepackage{graphicx}
\usepackage{slashed}
\begin{document}

\title{Neutrino masses in the minimal gauged $(B-L)$ supersymmetry}
\author{Yu-Li Yan\footnote{yychanghe@hbu.edu.cn}, Tai-Fu
Feng\footnote{fengtf@hbu.edu.cn}, Jin-Lei Yang, Hai-Bin Zhang\footnote{hbzhang@hbu.edu.cn}, Shu-Min
Zhao, Rong-Fei Zhu}

\affiliation{Department of Physics, Hebei University, Baoding, 071002, China}

\begin{abstract}

We present the radiative corrections to neutrino masses in a minimal supersymmetric extension of the standard model
 with local $U(1)_{B-L}$ symmetry. At tree level, three tiny active neutrinos and two nearly massless sterile neutrinos can be obtained through the seesaw mechanism. Considering the one-loop corrections to the neutrino masses, the numerical results indicate that two sterile neutrinos obtain ${\rm KeV}$ masses and the small active-sterile neutrino mixing angles. The lighter sterile neutrino is a very interesting dark matter candidate in cosmology. Meanwhile the active neutrinos mixing angles and mass squared differences agree with present experimental data.

\end{abstract}


\maketitle

\section{Introduction\label{sec1}}

The discovery of Higgs boson on the Large Hadron Collider (LHC) \cite{CMS,ATLAS} indicates that the Higgs mechanism to break electroweak symmetry has an experimental cornerstone now. The experiments of atmospheric and solar neutrino oscillation give the neutrino data at least three types of neutrinos which have sub-$\rm{eV}$ masses, but the standard model (SM) of particle physics cannot account for the origin of these tiny masses naturally.

Three flavor neutrinos are mixed into massive neutrinos $\nu_{1,2,3}$ during their flight, and the mixings are described by the Pontecorvo-Maki-Nakagawa-Sakata matrix $U_{PMNS}$ \cite{neutrino-oscillations,neutrino-oscillations1}. At one standard deviation, a global fitting from the updated neutrino oscillation experimental data gives the differences of mass squared and mixing angles as \cite{neutrino-number}
\begin{eqnarray}
&&\Delta m_{\odot}^2=7.54_{-0.22}^{+0.26}\times 10^{-5}\;{\rm eV}^2\;,\nonumber\\
&&\Delta m_{A}^2(\mathrm{NO})=2.43_{-0.06}^{+0.06}\times 10^{-3}\; {\rm eV}^2\nonumber\;, \qquad\Delta m_{A}^2(\mathrm{IO})=2.38_{-0.06}^{+0.06}\times 10^{-3}\; {\rm eV}^2\;,\nonumber\\
&&\sin^2\theta_{12}=0.308\pm 0.0017\;,\nonumber\\
&&\sin^2\theta_{23}(\mathrm{NO})=0.437_{-0.023}^{+0.033}\;,  \quad\qquad\quad\quad\sin^2\theta_{23}(\mathrm{IO})=0.455_{-0.031}^{+0.039}\;,\nonumber\\
&&\sin^2\theta_{13}(\mathrm{NO})=0.0234_{-0.0019}^{+0.0020}\;,  \qquad\quad\quad\sin^2\theta_{13}(\mathrm{IO})=0.0240_{-0.0022}^{+0.0019}\;.
\label{neutrino-oscillations2}
\end{eqnarray}
The values correspond to neutrino mass spectrum with normal ordering (NO) or inverted ordering (IO). To account for the neutrino oscillation data in Eq. (1), a theory beyond the SM is necessary.

The supersymmetric extension of the SM is a rather popular choice. The discrete symmetry R-parity is defined through $R=(-1)^{3(B-L)+2S}$, where $B$, $L$, and $S$ are baryon number, lepton number, and the spin of the particle, respectively \cite{R-parity}. In the minimal supersymmetry extension of SM (MSSM) with local $U(1)_{B-L}$ symmetry, the nonzero vacuum expectation values (VEVs) of the right-handed sneutrinos evoke the $(B-L)$ symmetry and R-parity spontaneously broken simultaneously \cite{Perez1,Perez2,Perez3,Perez4,Perez5,Perez6}.
At tree level, the MSSM with local $U(1)_{B-L}$ symmetry can generate three active neutrinos to interpret the neutrino oscillation through the seesaw mechanism; meanwhile, the model predicts that there are two sterile neutrinos. Nevertheless, two sterile neutrinos have far below $\rm{eV}$ masses at tree level \cite{Perez4,Perez5,Perez6,Feng}. Sterile neutrinos with $\rm{KeV}$ scale masses are a well-motivated dark matter candidate for two reasons. First, fermionic dark matter cannot have an arbitrarily small mass, since in dense regions it cannot be packed within an infinitely small volume for the Pauli principle. Second, sterile neutrinos have a small mixing with the active neutrinos which would enable a dark matter particle to decay into an active neutrino and a photon \cite{DM}. The Tremaine-Gunn bound indicates that a sterile neutrino mass must be greater than about $0.4\;\rm{KeV}$
\cite{TG,DM}. A strong bound on a sterile neutrino mass and mixing angle comes from the nondetection results of the monoenergetic X-rays by the decay of sterile neutrino \cite{DM}.
 Recently two groups reported evidence for a $3.55\;\rm{KeV}$ emission line which could be from the decay of a $7.1\;\rm{KeV}$ sterile neutrino with $\sin^2(2\theta)\sim10^{-10}$ or $10^{-11}$ \cite{sterile neutrino1,sterile neutrino2}, which is just below the previous X-ray bound. This observation is being fiercely discussed \cite{sterile neutrino3,sterile neutrino4,sterile neutrino5,sterile neutrino6}.

In this work, we investigate the origin of neutrino masses in the minimal gauged $(B-L)$ supersymmetry. There are five light neutrinos (three light active and two almost massless sterile neutrinos) at tree level, which agrees with the results in Refs.\cite{Perez4,Perez5,Perez6,Feng}.
The one-loop corrections to the light neutrinos are important to account for relevant experimental data \cite{loopSSM,loopzhao}. In this article we present the one-loop radiative corrections to neutrino masses and relevant mixing matrix in the MSSM with local $U(1)_{B-L}$ symmetry. The numerical results indicate that there is parameter space to give two sterile neutrinos $\rm{KeV}$ masses and the small active-sterile neutrino mixing angles. The lighter sterile neutrino is a very interesting dark matter candidate in cosmology. Meanwhile, the mass squared differences and mixing angles of active neutrinos coincide with the experimental data from the solar and atmospheric oscillations \cite{neutrino-number}.

Our presentation is organized as follows. In Sec. \ref{sec2}, we briefly summarize the main ingredients of the MSSM with local $U(1)_{B-L}$ symmetry and then present the mass matrix for neutralinos and neutrinos at tree level in Sec. \ref{sec3}. In Sec. \ref{sec4}, we analyze one-loop radiative corrections to the mass matrix. The numerical analysis for two possibilities on the neutrino mass spectrum (NO and IO) is given in Sec. \ref{sec5}, and Sec. \ref{sec6} gives a summary.

\section{The supersymmetric model with local $U(1)_{B-L}$ symmetry\label{sec2}}

When $U(1)_{B-L}$ is a local gauge symmetry, one can enlarge the local gauge group of the MSSM to $SU(3)_{C}\otimes SU(2)_{L}\otimes U(1)_{Y} \otimes U(1)_{B-L}$. In the model proposed in Refs. \cite{Perez4,Perez5,Perez6}, three exotic superfields for right-handed neutrinos are $\hat{N}_{i}^c\sim(1,1,0,1)$. Meanwhile, quantum numbers of the matter chiral superfields for quarks and leptons are given by
\begin{eqnarray}
&&\hat{Q}_{I}=\left(\begin{array}{c}\hat{U}_{I}\\
\hat{D}_{I}\end{array}\right)\sim\left(3,2,{1\over3},{1\over3}\right),\quad
\hat{L}_{I}=\left(\begin{array}{c}\hat{\nu}_{I}\\
\hat{E}_{I}\end{array}\right)\sim(1,2,-1,-1),\nonumber\\
&&\hat{U}_{I}^c\sim\left(3,1,{-{4\over3}},{-{1\over3}}\right),\quad
\hat{D}_{I}^c\sim\left(3,1,{2\over3},{-{1\over3}}\right),\quad
\hat{E}_{I}^c\sim(1,1,2,1),
\label{quantum-number1}
\end{eqnarray}
with $I=1,2,3$ denoting the index of generation. In addition, the quantum numbers of two Higgs doublets are assigned as
\begin{eqnarray}
&&\hat{H}_{u}=\left(\begin{array}{c}\hat{H}_{u}^+\\
\hat{H}_{u}^0\end{array}\right)\sim(1,2,1,0),\quad
\hat{H}_{d}=\left(\begin{array}{c}\hat{H}_{d}^0\\
\hat{H}_{d}^-\end{array}\right)\sim(1,2,-1,0).
\label{quantum-number2}
\end{eqnarray}
The superpotential of the MSSM with local $U(1)_{B-L}$ symmetry is written as \cite{Feng,Feng2,Feng3,Feng4}
\begin{eqnarray}
&&{\cal W}={\cal W}_{MSSM}+(Y_{N})_{IJ}\hat{H}_{u}^Ti\sigma_2\hat{L}_I\hat{N}_{J}^c,
\label{potential}
\end{eqnarray}
with ${\cal W}_{MSSM}$ denoting the superpotential of the MSSM. Correspondingly, the soft breaking terms of the MSSM with local $U(1)_{B-L}$ symmetry are generally given as
\begin{eqnarray}
&&{\cal L}_{soft}={\cal L}_{soft}^{MSSM}-(m_{\tilde{N}^c}^2)_{IJ}\tilde{N}_I^{c\ast}\tilde{N}_J^c-
(m_{BL}\lambda_{BL}\lambda_{BL}+m_{BBL}\lambda_{B}\lambda_{BL}+H.c.)\nonumber\\
&&\qquad\qquad+\{(A_N)_{IJ}H_u^Ti\sigma_2\tilde{L}_I\tilde{N}_J^c+H.c.\}.
\label{L}
\end{eqnarray}
In this formula ${\cal L}_{soft}^{MSSM}$ is the soft breaking terms of the MSSM, and $\lambda_{BL}$ denotes the gaugino of $U(1)_{B-L}$. After the $SU(2)_{_L}$ doublets $H_u$, $H_d$, $\tilde{L}_I$, and $SU(2)_{_L}$ singlets, $\tilde{N}_I^c$ obtain the nonzero VEVs,
\begin{eqnarray}
&&H_u=\left(\begin{array}{c}H_u^+\\
{1\over\sqrt{2}}\Big(\upsilon_{u}+H_u^0+iP_u^0\Big)\end{array}\right),~~~~~~
H_d=\left(\begin{array}{c}{1\over\sqrt{2}}\Big(\upsilon_{d}+H_d^0+iP_d^0\Big)\\
H_d^-\end{array}\right)\nonumber,\\
&&\tilde{L}_I=\left(\begin{array}{c}{1\over\sqrt{2}}\Big(\upsilon_{L_I}+\tilde{\nu}_{L_I}+iP_{\tilde{L}_I}^0\Big)\\
\tilde{L}_I^-\end{array}\right),~~~~~~
\tilde{N}_I^c={1\over\sqrt{2}}\Big(\upsilon_{N_I}+\tilde{\nu}_{R_I}+iP_{\tilde{N}_I}^0\Big),
\label{VEVs}
\end{eqnarray}
the R-parity is broken spontaneously, and the local gauge symmetry $SU(2)_{_L}\otimes U(1)_{_Y} \otimes U(1)_{_{B-L}}$ breaks down to the electromagnetic symmetry $U(1)_e$. Then, the tree level masses of neutral and charged gauge bosons are, respectively, formulated as
\begin{eqnarray}
&&m_Z^2={1\over4}(g_1^2+g_2^2)\upsilon_{EW}^2,\qquad m_W^2={1\over4}g_2^2\upsilon_{EW}^2,\nonumber\\
&&m_{Z_{BL}}^2=g_{BL}^2(\upsilon_N^2+\upsilon_{EW}^2-\upsilon_{SM}^2),
\label{mz,mw,mzBL}
\end{eqnarray}
with abbreviations $\upsilon_{SM}^2=\upsilon_u^2+\upsilon_d^2$, $\upsilon_{EW}^2=\upsilon_u^2+\upsilon_d^2+\sum\limits_{\alpha=1}^3\upsilon_{L_\alpha}^2$ and $\upsilon_N^2=\sum\limits_{\alpha=1}^3\upsilon_{N_\alpha}^2$. In addition, $g_2$, $g_1$ and $g_{BL}$ denote the gauge couplings of $SU(2)_L$, $U(1)_Y$, and $U(1)_{B-L}$, respectively.

To satisfy present electroweak precision observations, we assume the mass of a neutral $U(1)_{B-L}$ gauge boson $m_{Z_{BL}}>1\rm TeV$, which implies $\upsilon_N>1\rm TeV$ when $g_{BL}<1$ ($m_{Z_{BL}}\simeq{g_{BL}\upsilon_N}$) \cite{Perez6}. After the electroweak symmetry is broken spontaneously, the couplings between the left-handed neutrinos and neutralinos are
$({1/\sqrt2}){\upsilon_{N_J}{(Y_N)_{IJ}}\psi_{L_I^1}\psi_{H_u^2}}-g_{BL}\upsilon_{L_I}\psi_{L_I^1}(i\lambda_{BL})
-(1/2){g_1}\upsilon_{L_I}\psi_{L_I^1}(i\lambda_B)+(1/2){g_2}\upsilon_{L_I}\psi_{L_I^1}(i\lambda_A^3)$. Because of the $\rm{TeV}$ scale seesaw suppression, the Yukawa couplings ${(Y_N)}_{IJ}$ and nonzero VEVs $v_{L_I}$ of left-handed sneutrino are sufficiently small for tiny active neutrino masses, $(Y_N)_{IJ}\leq10^{-6}$ and $v_{L_I}\leq10^{-3}\;\rm GeV$ \cite{Perez4,Perez6}. Ignoring the terms which are negligible and assuming that the $3\times3$ matrices $m_{\tilde{L}}^2$, $m_{\tilde{N}^c}^2$ are real, we simplify the minimization conditions as \cite{Feng}
\begin{eqnarray}
&&\upsilon_u\{\mu^2+m_{H_u}^2+{{g_1^2+g_2^2}\over8}(2\upsilon_u^2-\upsilon_{EW}^2)\}-B\mu{\upsilon_d}\simeq0\nonumber,\\
&&\upsilon_d\{\mu^2+m_{H_d}^2+{{g_1^2+g_2^2}\over8}(2\upsilon_u^2-\upsilon_{EW}^2)\}-B\mu{\upsilon_u}\simeq0\nonumber,\\
&&\sum_{\alpha=1}^3(m_{\tilde{L}}^2)_{I\alpha}\upsilon_{L_\alpha}+
{\upsilon_u\over\sqrt2}\sum_{\alpha=1}^3(A_N)_{I\alpha}\upsilon_{N_\alpha}-{\mu\upsilon_d\over\sqrt2}\zeta_I\nonumber\\
&&\qquad-\upsilon_{L_I}\{{{g_1^2+g_2^2}\over8}(2\upsilon_u^2-\upsilon_{EW}^2)+{m_{Z_{BL}}^2\over2}\}\simeq0\nonumber,\\
&&\sum_{\alpha=1}^3(m_{\tilde{N}^c}^2)_{I\alpha}\upsilon_{N_\alpha}+{m_{Z_{BL}}^2\over2}\upsilon_{N_I}\simeq0,
\label{the minimization conditions}
\end{eqnarray}
where $\zeta_I=\sum\limits_{\alpha=1}^3(Y_N)_{I\alpha}\upsilon_{N_\alpha}$. Note that the first two minimization conditions for $H_u^0$ and $H_d^0$ are not greatly altered from those in the MSSM, the third condition originates from the linear terms of $\upsilon_{L_I}$, and the last equation implies that the vector $(\upsilon_{N_1},\upsilon_{N_2},\upsilon_{N_3})$ is an eigenvector of $3\times3$ mass squared matrix $m_{\tilde{N}^c}^2$ with eigenvalue ${-m_{Z_{BL}}^2/2}$. Considering the last minimization condition in Eq. (8), we formulate the symmetric $3\times3$ matrix as
\begin{eqnarray}
&&m_{\tilde{N}^c}^2=\left(\begin{array}{ccc}
\xi_{\tilde{N}_1^c}^2-{m_{Z_{BL}}^2/2} & 0 & -{\upsilon_{N_1}\over{\upsilon_{N_3}}}\xi_{\tilde{N}_1^c}^2\\
0 & \xi_{\tilde{N}_2^c}^2-{m_{Z_{BL}}^2/2} & -{\upsilon_{N_2}\over{\upsilon_{N_3}}}\xi_{\tilde{N}_2^c}^2\\
-{\upsilon_{N_1}\over{\upsilon_{N_3}}}\xi_{\tilde{N}_1^c}^2 &  -{\upsilon_{N_2}\over{\upsilon_{N_3}}}\xi_{\tilde{N}_2^c}^2 & {{\xi_{\tilde{N}_1^c}^2{\upsilon_{N_1}}+\xi_{\tilde{N}_2^c}^2{\upsilon_{N_2}}}\over{\upsilon_{N_3}}}-{m_{Z_{BL}}^2/2} \end{array}\right),
\label{mN2}
\end{eqnarray}
with $\xi_{\tilde{N}_1^c}^2=(m_{\tilde{N}^c}^2)_{11}+{m_{Z_{BL}}^2/2}$, $\xi_{\tilde{N}_2^c}^2=(m_{\tilde{N}^c}^2)_{22}+{m_{Z_{BL}}^2/2}$. This is the mixing matrix of the right-handed sneutrinos, the reasons for choosing it are shown in Appendix A.

\section{The mass matrix for neutralinos and neutrinos at tree level\label{sec3}}

In the MSSM with local $U(1)_{B-L}$ symmetry, the nonzero VEVs of left- and right-handed sneutrinos induce the mixing between neutralinos (charginos) and neutrinos (charged leptons). In the basis $\Psi^{0T}=(\nu_{L_I},N_J^c,i\lambda_{BL},i\lambda_B,i\lambda_A^3,\psi_{H_d}^1,\psi_{H_u}^2)$, we can obtain the neutral fermion mass terms in the Lagrangian
\begin{eqnarray}
&&{\cal L}_{neutral}^{mass}={1\over2}\Psi^{0T}M_N\Psi^{0}+H.c.,
\label{L}
\end{eqnarray}
where the mass matrix for neutralinos and neutrinos $M_N$ is given by
\begin{eqnarray}
&&M_N=\left(\begin{array}{ccc}
0_{3\times3} & ({\cal A}_N^{(1)})_{3\times4} & ({\cal A}_N^{(2)})_{3\times4}\\
({\cal A}_N^{(1)T})_{4\times3} & ({\cal M}_N^{(0)})_{4\times4} & ({\cal A}_N^{(3)})_{4\times4}\\
({\cal A}_N^{(2)T})_{4\times3} & ({\cal A}_N^{(3)T})_{4\times4} & ({\cal M}_N)_{4\times4}
\end{array}\right),
\label{MN}
\end{eqnarray}
where ${\cal M}_N$ denotes the $4\times4$ mass matrix for neutralinos in the MSSM. The concrete expressions for ${\cal M}_N^{(0)}$, ${\cal A}_N^{(1)}$, ${\cal A}_N^{(2)}$, and ${\cal A}_N^{(3)}$ are
\begin{eqnarray}
&&{\cal M}_N^{(0)}=\left(\begin{array}{cc}
0_{3\times3} & (g_{BL}\upsilon_{N_J})_{3\times1}\\
(g_{BL}\upsilon_{N_{J'}})_{1\times3} & 2m_{BL}
\end{array}\right),\nonumber
\label{MN0}
\end{eqnarray}
\begin{eqnarray}
&&\quad\qquad{\cal A}_N^{(1)}=\left(\begin{array}{cc}
({{{\upsilon_u}\over{\sqrt2}}}(Y_N)_{IJ'})_{3\times3}\quad & (-g_{BL}\upsilon_{L_I})_{3\times1}
\end{array}\right),\nonumber
\label{AN1}
\end{eqnarray}
\begin{eqnarray}
&&\quad\qquad\qquad\qquad\qquad{\cal A}_N^{(2)}=\left(\begin{array}{cccc}
(-{g_1\over2}\upsilon_{L_I})_{3\times1}\quad & ({g_2\over2}\upsilon_{L_I})_{3\times1}\quad & 0_{3\times1}\quad & ({\zeta_I\over{\sqrt2}})_{3\times1}
\end{array}\right),\nonumber
\label{AN2}
\end{eqnarray}
\begin{eqnarray}
&&\quad\qquad\qquad\qquad\qquad{\cal A}_N^{(3)}=\left(\begin{array}{cccc}
0_{3\times1} & 0_{3\times1} & 0_{3\times1} & ({1\over{\sqrt2}}\sum\limits_{\alpha=1}^3\upsilon_{L_\alpha}(Y_N)_{\alpha J})_{3\times1}\\
m_{BBL}  &  0 & 0   &   0
\end{array}\right),\qquad
\label{AN3}
\end{eqnarray}
with the row indices of matrix $I,J=1,2,3$ and the column indices of matrix $I',J'=1,2,3$, respectively. The eigenvalues of the $4\times4$ mass matrix ${\cal M}_N^{(0)}$
are given as
\begin{eqnarray}
&&m_{N_1}=m_{N_2}=0,\qquad m_{N_3}={m_{BL}-\Delta_{BL}},\qquad m_{N_4}={m_{BL}+\Delta_{BL}},
\label{MN0}
\end{eqnarray}
where $\Delta_{BL}=\sqrt{m_{BL}^2+g_{BL}^2\upsilon_{N}^2}$, and then we can obtain $m_{N_{3,4}}$ about $\rm TeV$ region for $m_{Z_{BL}}>1\rm TeV$ ($m_{Z_{BL}}\simeq{g_{BL}\upsilon_N}$). The matrix ${\cal M}_N^{(0)}$ has four eigenvalues which are zero order approximations of the $U(1)_{B-L}$ gaugino and three right-handed neutrinos masses. However, there are only two nonzero eigenvalues, and the other two eigenvalues are zero.

Defining the $4\times4$ orthogonal matrix
\begin{eqnarray}
&&U_N=\left(\begin{array}{cccc}
 -{\upsilon_{N_3}\over{\sqrt{\upsilon_{N_1}^2+\upsilon_{N_3}^2}}} &  -{\upsilon_{N_1}\upsilon_{N_2}\over{\upsilon_N\sqrt{\upsilon_{N_1}^2+\upsilon_{N_3}^2}}} &  -{{g_{BL}\upsilon_{N_1}}\over{\sqrt2\Delta_{BL}\eta_{BL}^-}} & {g_{BL}\upsilon_{N_1}}\over{\sqrt2\Delta_{BL}\eta_{BL}^+}\\
 0 & \sqrt{\upsilon_{N_1}^2+\upsilon_{N_3}^2}\over{\upsilon_N} &  -{{g_{BL}\upsilon_{N_2}}\over{\sqrt2\Delta_{BL}\eta_{BL}^-}} & {g_{BL}\upsilon_{N_2}}\over{\sqrt2\Delta_{BL}\eta_{BL}^+}\\
 \upsilon_{N_1}\over{\sqrt{\upsilon_{N_1}^2+\upsilon_{N_3}^2}}&  -{\upsilon_{N_3}\upsilon_{N_2}\over{\upsilon_N\sqrt{\upsilon_{N_1}^2+\upsilon_{N_3}^2}}} &  -{{g_{BL}\upsilon_{N_3}}\over{\sqrt2\Delta_{BL}\eta_{BL}^-}} & {g_{BL}\upsilon_{N_3}}\over{\sqrt2\Delta_{BL}\eta_{BL}^+}\\
0 & 0 & {1\over{\sqrt2}}\eta_{BL}^- & {1\over{\sqrt2}}\eta_{BL}^+
\end{array}\right),
\label{UN}
\end{eqnarray}
one obtains
\begin{eqnarray}
&&M=\left(\begin{array}{ccc}
1_{3\times3} & 0_{3\times4} & 0_{3\times4}\\
0_{4\times3} & (U_N^T)_{4\times4} & 0_{4\times4}\\
0_{4\times3} & 0_{4\times4} & 1_{4\times4}
\end{array}\right)\cdot{M_N}\cdot
\left(\begin{array}{ccc}
1_{3\times3} & 0_{3\times4} & 0_{3\times4}\\
0_{4\times3} & (U_N)_{4\times4} & 0_{4\times4}\\
0_{4\times3} & 0_{4\times4} & 1_{4\times4}
\end{array}\right)\nonumber\\
&&\quad=\left(\begin{array}{ccc}
0_{3\times3} & ({\cal A}_N^{(1)}U_N)_{3\times4} & ({\cal A}_N^{(2)})_{3\times4}\\
(U_N^T{\cal A}_N^{(1)T})_{4\times3} & (U_N^T{\cal M}_N^{(0)}U_N)_{4\times4} & (U_N^T{\cal A}_N^{(3)})_{4\times4}\\
({\cal A}_N^{(2)T})_{4\times3} & ({\cal A}_N^{(3)T}U_N)_{4\times4} & ({\cal M}_N)_{4\times4}
\end{array}\right)\nonumber\\
&&\quad=\left(\begin{array}{cc}
(m_\nu)_{5\times5} & (m_D)_{5\times6}\\
(m_D^T)_{6\times5} & ({\cal M})_{6\times6}
\end{array}\right),
\label{UNTMUN}
\end{eqnarray}
where $\eta_{BL}^\pm=\sqrt{1\pm{m_{BL}\over{\Delta_{BL}}}}$,  and  $U_N^T{\cal M}_N^{(0)}U_N=diag(0,\quad0,
\quad{m_{BL}-\Delta_{BL}},\quad{m_{BL}+\Delta_{BL}}),\;$ respectively.

Using Eqs. (11) and (14), we formulate the submatrices in Eq. (15), respectively, as
\begin{eqnarray}
&&m_\nu=\left(\begin{array}{ccc}
0_{3\times3} &  \delta_{i3} & \delta_{i2}\\
\delta_{i3} & 0 & 0\\
\delta_{i2} & 0 & 0
\end{array}\right),\qquad\qquad
m_D=\left(\begin{array}{cccccc}
-\delta_i^- & \delta_i^+  & -{g_1\over2}\upsilon_{L_i} & {g_2\over2}\upsilon_{L_i} & 0_{3\times1} & {1\over{\sqrt2}}\zeta_i\\
0 & 0 & 0 & 0 & 0 & \varepsilon_{13}\\
0 & 0 & 0 & 0 & 0 & \varepsilon_{12}
\end{array}\right),\nonumber\\
&&{\cal M}=\left(\begin{array}{cccccc}
m_{BL}-\Delta_{BL} & 0 & {1\over\sqrt2}\eta^-m_{BBL} & 0 & 0 & -\varepsilon_{-}\\
0 & m_{BL}+\Delta_{BL} & {1\over\sqrt2}\eta^+m_{BBL} & 0 & 0 & \varepsilon_{+}\\
{1\over\sqrt2}\eta^-m_{BBL} & {1\over\sqrt2}\eta^+m_{BBL} & M_1 & 0 & -{g_1\over2}\upsilon_d & {g_1\over2}\upsilon_u\\
0 & 0 & 0 & M_2 & {g_2\over2}\upsilon_d & -{g_2\over2}\upsilon_u\\
0 & 0 & -{g_1\over2}\upsilon_d & {g_2\over2}\upsilon_d & 0 & -\mu\\
-\varepsilon_{-} & \varepsilon_{+} & {g_1\over2}\upsilon_u &  -{g_2\over2}\upsilon_u & -\mu & 0\\
\end{array}\right),
\label{MmDmv}
\end{eqnarray}
where the abbreviations are
\begin{eqnarray}
&&\varepsilon_N^2=\sum\limits_{\alpha,\beta=1}^3\upsilon_{L_\alpha}(Y_N)_{\alpha\beta}\upsilon_{N_\beta},\qquad\qquad\qquad
\varepsilon_\pm={{g_{_{BL}}\varepsilon_N^2}\over{2\Delta_{BL}\eta_{_{BL}}^\pm}},\nonumber\\
&&{\delta_i^\pm}={{g_{_{BL}}\upsilon_u}\over{2\Delta_{_{BL}}\eta_{_{BL}}^\pm}}\zeta_i\mp{1\over{\sqrt2}}g_{BL}\eta_{_{BL}}^\pm\upsilon_{L_i},\nonumber\\
&&\delta_{i2}={\upsilon_u\over{\sqrt{2(\upsilon_{N_1}^2+\upsilon_{N_3}^2)}}}
[-(Y_N)_{i1}\upsilon_{N_1}\upsilon_{N_2}+(Y_N)_{i2}(\upsilon_{N_1}^2+\upsilon_{N_3}^2)
-(Y_N)_{i3}\upsilon_{N_2}\upsilon_{N_3}],\nonumber\\
&&\delta_{i3}={\upsilon_u\over{\upsilon_N\sqrt{2(\upsilon_{N_1}^2+\upsilon_{N_3}^2)}}}
[-(Y_N)_{i1}\upsilon_{N_3}+(Y_N)_{i3}\upsilon_{N_1}],\nonumber\\
&&\varepsilon_{12}={1\over{\upsilon_u}}\sum\limits_{\alpha=1}^3\upsilon_{L_\alpha}\delta_{\alpha2},\qquad\qquad\quad
\varepsilon_{13}={1\over{\upsilon_u}}\sum\limits_{\alpha=1}^3\upsilon_{L_\alpha}\delta_{\alpha3},
\label{abbreviations}
\end{eqnarray}
with the indices $i=1,2,3$. The abbreviations are suppressed by the tiny $(Y_N)_{ij}$ and $\upsilon_{L_i}$, and they are very small.

Defining the $11\times11$ approximated orthogonal transformation matrix ${\cal Z}_N$ \cite{Feng}
\begin{eqnarray}
{{\cal Z}_N}=\left(\begin{array}{cc}
[1-{1\over2}m_D\cdot{\cal M}^{-2}\cdot{m_D^T}]_{5\times5} &  [m_D\cdot{\cal M}^{-1}+m_\nu\cdot{m_D\cdot{\cal M}^{-2}}]_{5\times6}\\
-[{\cal M}^{-1}\cdot{m_D^T}+{\cal M}^{-2}\cdot{m_D^T}\cdot{m_\nu}]_{6\times5} & [1-{1\over2}{\cal M}^{-1}\cdot{m_D^T}\cdot{m_D\cdot{\cal M}^{-1}}]_{6\times6}
\end{array}\right),
\label{Zn}
\end{eqnarray}
via the seesaw mechanism, we finally write the effective mass matrix for five light neutrinos (three active and two sterile) as
\begin{eqnarray}
&&{m_{eff}}\simeq{m_\nu-m_D\cdot{\cal M}^{-1}\cdot{m_D^T}-{1\over2}m_\nu\cdot{m_D\cdot{\cal M}^{-2}}\cdot{m_D^T}-{1\over2}m_D\cdot{\cal M}^{-2}\cdot{m_D^T}\cdot{m_\nu}}\nonumber\\
&&\qquad\;\simeq\left(\begin{array}{cc}
[M_\nu^{LL}]_{3\times3} & [M_\nu^{LR}]_{3\times2}\\
{[M_\nu^{LR,T}]_{2\times3}} & [M_\nu^{RR}]_{2\times2}
\end{array}\right),
\label{meff}
\end{eqnarray}
where $M_\nu^{LL}$ is the three active Majorana mass matrices, $M_\nu^{LR}$ is the Dirac mixing mass matrix and $M_\nu^{RR}$ is the two sterile Majorana mass matrices. The concrete expression of $M_\nu^{LL}$, $M_\nu^{LR}$, and $M_\nu^{RR}$ can be found in Appendix B. From Eqs. (B2) and (17) the elements of $M_\nu^{RR}$ are related to the square of tiny $(Y_N)_{ij}\upsilon_{L_i}$; therefore, they are almost zero. The two sterile neutrinos are almost massless at tree level.

In order to accommodate naturally the experimental data on neutrino oscillation and $Z$ invisible decay width in this framework, one can find that only one possibility $M_\nu^{LL}\gg{M_\nu^{LR}},\;{M_\nu^{RR}}$ is reasonable \cite{Perez4}. In fact, from Eq. (B2) this point implies
\begin{eqnarray}
\delta_{i2},\delta_{i3}\ll{{\upsilon_{L_i}^2\over{\Lambda_\upsilon}}+{\zeta_i^2\over{\Lambda_\zeta}}+
{{2\upsilon_{L_i}\zeta_i}\over{\Lambda_{\upsilon\zeta}}}}.
\label{di2}
\end{eqnarray}

To guarantee the decoupling of two tiny sterile neutrinos from the active neutrinos, we choose the Yukawa coupling for right-handed neutrinos as
\begin{eqnarray}
Y_N={1\over{\upsilon_N}}\left(\begin{array}{ccc}
\upsilon_{N_1}Y_1   & \upsilon_{N_2}Y_1 & \upsilon_{N_3}Y_1 \\
\upsilon_{N_1}Y_2   & \upsilon_{N_2}Y_2 & \upsilon_{N_3}Y_2 \\
\upsilon_{N_1}Y_3   & \upsilon_{N_2}Y_3 & \upsilon_{N_3}Y_3
\end{array}\right)={1\over{\upsilon_N}}\left(\begin{array}{ccc}
Y_1 & 0 & 0 \\
0 & Y_2 & 0 \\
0 & 0 & Y_3
\end{array}\right)\left(\begin{array}{ccc}
\upsilon_{N_1}  & \upsilon_{N_2} & \upsilon_{N_3} \\
\upsilon_{N_1}  & \upsilon_{N_2} & \upsilon_{N_3} \\
\upsilon_{N_1}  & \upsilon_{N_2} & \upsilon_{N_3}
\end{array}\right),
\label{YN}
\end{eqnarray}
then we get $\zeta_i=Y_i\upsilon_N$,  $\delta_{i2}=\delta_{i3}=0$, $(i=1,2,3)$.

Only including the tree level contributions to the light neutrino mass matrix in Eq. (19), we diagonalize the effective neutrino mass matrix $m_{eff}$ and then obtain three light left-handed neutrinos and two nearly massless sterile neutrinos \cite{Perez4,Perez5,Perez6,Feng}. Recently, it has been shown in Refs. \cite{DM,sterile neutrino1,sterile neutrino2} that sterile neutrinos with $\rm{KeV}$ scale masses are interesting dark matter candidates in the Universe. The one-loop radiative corrections are important, especially for the light neutrinos \cite{Arise1,Arise2}. We consider the one-loop radiative corrections to the mass matrix of the neutrinos in Eq. (15) and expect that two sterile neutrinos acquire their physical masses at $\rm KeV$ level in the following.

\section{The radiative corrections on masses of neutrinos\label{sec4}}
\subsection{The radiative corrections on masses of sterile neutrinos}

In this model, there is a large mixing between three right-handed neutrinos and a $(B-L)$ gaugino. At leading order, this mixing induces two heavy Majorana fermions with masses about
 $\rm TeV$ scale and two light sterile neutrinos which acquire their tiny masses by a seesaw mechanism. Here, we consider one-loop radiative corrections to the masses of two light sterile neutrinos. From interactions of the gauge and matter multiplets $ig{\sqrt2}T_{ij}^2(\lambda^a\psi_jA_i^\ast-\bar{\lambda}^a\bar{\psi}_iA_j)$, $-gT_{ij}^aV_\mu^a\bar{\psi}_i\bar{\sigma}^\mu\psi_j$ \cite{ MSSM}, we can obtain the couplings involving sterile neutrinos.
In the Majorana case, the most general form for $N_\alpha\rightarrow{N_\beta}$ transition reads
\begin{eqnarray}
&&\Sigma_{\alpha\beta}(\slashed p)=\Sigma_{\alpha\beta}^L(p^2) {\slashed p}P_L+\Sigma_{\alpha\beta}^{L\ast}(p^2){\slashed p}P_R+\Sigma_{\alpha\beta}^M(p^2)P_L+\Sigma_{\alpha\beta}^{M\ast}(p^2)P_R.
\label{PLPR}
\end{eqnarray}
The invariance of \textit{CPT} transformation requires
\begin{eqnarray}
&&\Sigma_{\alpha\beta}^L(p^2)=\Sigma_{\beta\alpha}^{L\ast}(p^2),\nonumber\\
&&\Sigma_{\alpha\beta}^M(p^2)=\Sigma_{\beta\alpha}^{M\ast}(p^2).
\label{PLPR}
\end{eqnarray}

The radiative corrections from real and image components of scalar right-handed neutrinos are
\begin{eqnarray}
&&\Sigma_{\alpha\beta}^{L(1)}(p^2)={2g_{BL}^2\over{(4\pi)^2}}\sum_{\delta=1}^4\sum_{i,j}^3
(U_N)_{i\alpha}^\ast(U_N)_{4\delta}^\ast(U_N)_{j\delta}(U_N)_{4\beta}\nonumber\\
&&\hspace{2.0cm}\times[\sum_{a=1}^3(U_{\tilde{N}_E})_{ia}(U_{\tilde{N}_E})_{ja}B_1(p^2,m_{N_\delta}^2,m_{H_{N_a}}^2)\nonumber\\
&&\hspace{2.0cm}+\sum_{a=1}^3(U_{\tilde{N}_O})_{ia}(U_{\tilde{N}_O})_{ja}B_1(p^2,m_{N_\delta}^2,m_{P_{N_a}}^2)],\nonumber\\
&&\Sigma_{\alpha\beta}^{M(1)}(p^2)={2g_{BL}^2\over{(4\pi)^2}}\sum_{\delta=1}^4\sum_{i,j}^3
m_{N_\delta}(U_N)_{i\alpha}(U_N)_{4\delta}(U_N)_{j\delta}(U_N)_{4\beta}\nonumber\\
&&\hspace{2.0cm}\times[\sum_{a=1}^3(U_{\tilde{N}_E})_{ia}(U_{\tilde{N}_E})_{ja}B_0(p^2,m_{N_\delta}^2,m_{H_{N_a}}^2)\nonumber\\
&&\hspace{2.0cm}-\sum_{a=1}^3(U_{\tilde{N}_O})_{ia}(U_{\tilde{N}_O})_{ja}B_0(p^2,m_{N_\delta}^2,m_{P_{N_a}}^2)].
\label{N-N1}
\end{eqnarray}
where $U_{\tilde{N}_O}$, $U_{\tilde{N}_E}$ can be found in Appendix B, and $U_N$ is given in Eq. (14).

In a similar way, the radiative corrections from a $(B-L)$ gauge boson are written as
\begin{eqnarray}
&&\Sigma_{\alpha\beta}^{L(2)}(p^2)={g_{BL}^2\over{2(4\pi)^2}}\sum_{\delta=1}^4\sum_{i,j}^3
(U_N)_{i\alpha}^\ast(U_N)_{i\delta}(U_N)_{j\delta}^\ast(U_N)_{j\beta}(B_1(p^2,m_{N_\delta}^2,m_{Z_{BL}}^2)-{1\over2}),\nonumber\\
&&\Sigma_{\alpha\beta}^{M(2)}(p^2)={g_{BL}^2\over{(4\pi)^2}}\sum_{\delta=1}^4\sum_{i,j}^3
m_{N_\delta}(U_N)_{i\alpha}(U_N)_{i\delta}^\ast(U_N)_{j\delta}^\ast(U_N)_{j\beta}(B_0(p^2,m_{N_\delta}^2,m_{Z_{BL}}^2)-{1\over2}).
\label{N-N2}
\end{eqnarray}
$B_0$ and $B_1$ are two-point scalar functions. The definitions are
\begin{eqnarray}
&&B_0(p^2,m_1^2,m_2^2)={(2\pi\Lambda)^{2\varepsilon}\over{i\pi^2}}\int{{d^Dq}\over{(q^2-m_1^2)((q-p)^2-m_2^2)}},\nonumber\\
&&{p_\mu}B_1(p^2,m_1^2,m_2^2)={(2\pi\Lambda)^{2\varepsilon}\over{i\pi^2}}\int{{d^Dqq_\mu}\over{(q^2-m_1^2)((q-p)^2-m_2^2)}},
\label{B0B1}
\end{eqnarray}
with $\varepsilon=2-{D\over2}$, where $\Lambda$ denotes the energy scale of new physics, and $\Lambda=2\;\rm{TeV}$ in our numerical analysis.

The generic expression for the right-handed Majorana neutrinos and the $(B-L)$ gaugino self-energy must be symmetric in its indices $\alpha,\beta$, and the result of one-loop corrections to the mass matrix in the modified dimensional reduction $(\overline{DR})$ scheme \cite{loopSSM} is written as
\begin{eqnarray}
&&(\Delta{\cal M}_N^{(0)})_{\alpha\beta}={1\over2}[{\cal{R}} (\hat{\Sigma}_{\alpha\beta}^M(m_{N_\alpha}^2))
+{\cal{R}} (\hat{\Sigma}_{\beta\alpha}^M(m_{N_\beta}^2))\nonumber\\
&&\hspace{2.6cm}-m_{N_{\alpha}}{\cal{R}}(\hat{\Sigma}_{\alpha\beta}^L(m_{N_\alpha}^2))
-m_{N_{\beta}}{\cal{R}}(\hat{\Sigma}_{\beta\alpha}^L(m_{N_\beta}^2))],
\label{DR}
\end{eqnarray}
where $\hat{\Sigma}$ denotes the renormalized self-energy in the $\overline{DR}$ scheme.

Using the concrete expression of $U_N$ in Eq. (14) ($(U_N)_{41}=(U_N)_{42}=0$) and the fact $m_{N_{1,2}}\ll{m_{N_{3,4}}}$ in Eq. (13), when $\alpha,\beta\leq2$, we find
\begin{eqnarray}
&&\hat{\Sigma}_{\alpha\beta}^L(p^2)=\hat{\Sigma}_{\alpha\beta}^{L(2)}(p^2),\qquad
\hat{\Sigma}_{\alpha\beta}^M(p^2)=\hat{\Sigma}_{\alpha\beta}^{M(2)}(p^2),
\label{M12}
\end{eqnarray}
then
\begin{eqnarray}
&&(\Delta{\cal M}_N^{(0)})_{\alpha\beta}={g_{BL}^2\over{2(4\pi)^2}}\sum_{\delta=3}^4\sum_{i,j}^3
m_{N_\delta}[{\cal{R}}((U_N)_{i\alpha}(U_N)_{i\delta}^\ast(U_N)_{j\delta}^\ast(U_N)_{j\beta})+(\alpha\leftrightarrow\beta)]\nonumber\\
&&\hspace{2.5cm}\times(\hat{B}_0(m_{N_\alpha}^2,m_{N_\delta}^2,m_{Z_{BL}}^2)-{1\over2}),
\label{MN12}
\end{eqnarray}

where two-point scalar functions $B_0$ and $B_1$ are renormalized in the $\overline{DR}$ scheme, denoted by $\hat{B}_0$ and $\hat{B}_1$, respectively. The corrections to the masses of sterile neutrino have nothing to do with small $(Y_N)_{ij}$ and $\upsilon_{L_i}$. As $\alpha=1,2$, $\beta=3,4$ and $\alpha,\beta=3,4$, the results of one-loop corrections to the mass matrix can be found in Appendix C.

\subsection{The radiative corrections on masses of three active neutrinos}

Generally the interaction between the left-handed neutrino and neutralinos, $(B-L)$ gaugino and charginos also induces the radiative corrections to the masses of left-handed neutrinos. However, these corrections are suppressed by the tiny Yukawa Couplings $(Y_N)_{ij}$ and VEVs $\upsilon_{L_i}$ of the left-handed sneutrino.  The results of one-loop radiative corrections to three active neutrinos \cite{active loop,active loop2} are
\begin{eqnarray}
&&(\Delta{\cal M}_L^{(0)})_{ij}={{\alpha_{EW}\delta({m_{\tilde{\nu}}^2})_{LL}^{ij}}\over{{4\pi}s_W^2c_W^2\Lambda^2}}
\sum_{\alpha=1}^4(c_W(U_{\chi^0})_{2\alpha}-s_W(U_{\chi^0})_{1\alpha}){m_{\chi_\alpha^0}}
\varrho_{1,1}(x_{\chi_\alpha^0},x_{\tilde{\nu}_{L_i}},x_{\tilde{\nu}_{L_j}})\nonumber\\
&&\hspace{2.2cm}+{{\alpha_{BL}\delta({m_{\tilde{\nu}}^2})_{LL}^{ij}}\over{2\pi\Lambda^2}}
\sum_{\alpha=3}^4(U_N)_{4\alpha}{m_{N_\alpha}}\varrho_{1,1}(x_{N_\alpha},x_{\tilde{\nu}_{L_i}},x_{\tilde{\nu}_{L_j}})\nonumber\\
&&\hspace{2.2cm}+{{\alpha_{EW}\delta({m_{\tilde{e}}^2})_{LL}^{ij}}\over{2\pi s_W^2\Lambda^2}}
\sum_{\alpha=1}^2(U_\pm)_{1\alpha}^2{m_{\chi_\alpha^\pm}}\varrho_{1,1}(x_{\chi_\alpha^\pm},x_{\tilde{e}_{L_i}},x_{\tilde{e}_{L_j}})\nonumber\\
&&\hspace{2.2cm}-{{\alpha_{EW}\mu^2\zeta_i\zeta_j}\over{{16\pi}s_W^2c_W^2{s_\beta^2}\Lambda^4}}
\sum_{\alpha=1}^4(c_W(U_{\chi^0})_{2\alpha}-s_W(U_{\chi^0})_{1\alpha}){m_{\chi_\alpha^0}}\nonumber\\
&&\hspace{2.2cm}\times\{\cos^2(\alpha-\beta)\varrho_{1,1}(x_{\chi_\alpha^0},x_{\tilde{\nu}_{L_i}},x_{\tilde{\nu}_{L_j}},x_h)\nonumber\\
&&\hspace{2.2cm}+\sin^2(\alpha-\beta)\varrho_{1,1}(x_{\chi_\alpha^0},x_{\tilde{\nu}_{L_i}},x_{\tilde{\nu}_{L_j}},x_H)
-\varrho_{1,1}(x_{\chi_\alpha^0},x_{\tilde{\nu}_{L_i}},x_{\tilde{\nu}_{L_j}},x_A)\}\nonumber\\
&&\hspace{2.2cm}-{{\alpha_{BL}\mu^2\zeta_i\zeta_j}\over{{4\pi}s_\beta^2\Lambda^4}}
\sum_{\alpha=3}^4(U_N)_{4\alpha}^2{m_{N_\alpha}}
\{\cos^2(\alpha-\beta)\varrho_{1,1}(x_{N_\alpha},x_{\tilde{\nu}_{L_i}},x_{\tilde{\nu}_{L_j}},x_h)\nonumber\\
&&\hspace{2.2cm}+\sin^2(\alpha-\beta)\varrho_{1,1}(x_{N_\alpha},x_{\tilde{\nu}_{L_i}},x_{\tilde{\nu}_{L_j}},x_H)
-\varrho_{1,1}(x_{N_\alpha},x_{\tilde{\nu}_{L_i}},x_{\tilde{\nu}_{L_j}},x_A)\}\nonumber\\
&&\hspace{2.2cm}-{{\alpha_{EW}\mu^2\zeta_i\zeta_j}\over{{4\pi}s_W^2s_\beta^2\Lambda^4}}
\sum_{\alpha=1}^2(U_\pm)_{1\alpha}{m_{\chi_\alpha^\pm}}
\varrho_{1,1}(x_{\chi_\alpha^\pm},x_{\tilde{e}_{L_i}},x_{\tilde{e}_{L_j}},x_{H_\pm}),
\label{ML1}
\end{eqnarray}
with $\alpha_{EW}={e^2/{4\pi}}$, $\alpha_{BL}={g_{BL}^2/{4\pi}}$ and $\delta({m_{\tilde{\nu}}^2})_{LL}^{ij}=({{g_1^2+g_2^2}\over4}+g_{BL}^2)\upsilon_{L_i}\upsilon_{L_j}+{1\over2}\zeta_i\zeta_j
+{1\over2}(Y_NY_N^T)_{ij}\upsilon_u^2$. Here, $U_{\chi^0}$ denotes the orthogonal matrix of a neutralino mass matrix, and $U_\pm$ denotes the orthogonal matrix of a chargino mass matrix in the MSSM. We also adopt the abbreviations $\tan\beta=\upsilon_u/\sqrt{\upsilon_d^2+\sum_{\alpha=1}^3\upsilon_{L_\alpha}^2}$, $\tan{2\alpha}={{m_A^2+m_Z^2}\over{{m_A^2-m_Z^2}}}\tan{2\beta}$, $s_\beta^2=\sin^2\beta$, $s_W^2=\sin^2{\theta_W}$, and  $c_W^2=\cos^2{\theta_W}$, where $\theta_W$ is the Weinberg angle. Here the functions $\varrho_{m,n}(x_1,x_2,\cdot\cdot\cdot,x_N)$ are defined by
\begin{eqnarray}
\varrho_{m,n}(x_1,x_2,\cdot\cdot\cdot,x_N)=\sum_{i=1}^N{{x_i^mln^nx_i}\over{\prod_{j\neq{i}}(x_i-x_j)}},
\label{p11}
\end{eqnarray}
with $x_i=m_i^2/{\Lambda^2}$.

Similarly, we derive the corrections from virtual sneutrino-neutralino to the mixing matrix ${\cal A}_N^{(2)}$ at a one-loop level as \cite{active loop,active loop2}
\begin{eqnarray}
&&\Delta{\cal A}_N^{(2)}=\sum_{k=1}^4(N_F^{(2)})_{ik}\{((s_\alpha(U_{\chi^0})_{3k}+c_\alpha(U_{\chi^0})_{4k})s_W,\;
-(s_\alpha(U_{\chi^0})_{3k}+c_\alpha(U_{\chi^0})_{4k})c_W,\;\nonumber\\
&& -(c_W(U_{\chi^0})_{2k}-s_W(U_{\chi^0})_{1k})s_\alpha,\;
-(c_W(U_{\chi^0})_{2k}-s_W(U_{\chi^0})_{1k})c_\alpha)\varrho_{1,1}(m_{\chi_k^0},x_h,x_{\tilde{\nu}_i})cos(\alpha-\beta)\nonumber\\
&&+(-(c_\alpha(U_{\chi^0})_{3k}+s_\alpha(U_{\chi^0})_{4k})s_W,\;
(c_\alpha(U_{\chi^0})_{3k}-s_\alpha(U_{\chi^0})_{4k})c_W,\;\nonumber\\
&& (c_W(U_{\chi^0})_{2k}-s_W(U_{\chi^0})_{1k})c_\alpha,\;
-(c_W(U_{\chi^0})_{2k}-s_W(U_{\chi^0})_{1k})s_\alpha)\varrho_{1,1}(m_{\chi_k^0},x_H,x_{\tilde{\nu}_i})sin(\alpha-\beta)\nonumber\\
&&+((-s_\beta(U_{\chi^0})_{3k}+c_\beta(U_{\chi^0})_{4k})s_W,\;
(s_\beta(U_{\chi^0})_{3k}-c_\beta(U_{\chi^0})_{4k})c_W,\;\nonumber\\
&& -(c_W(U_{\chi^0})_{2k}-s_W(U_{\chi^0})_{1k})c_\beta,\;
-(c_W(U_{\chi^0})_{2k}-s_W(U_{\chi^0})_{1k})s_\beta)\varrho_{1,1}(m_{\chi_k^0},x_A,x_{\tilde{\nu}_i})\},
\label{AN2}
\end{eqnarray}
with $s_\alpha^2=\sin^2\alpha$, $c_\alpha^2=\cos^2\alpha$, and $(N_F^{(2)})_{ik}={{\alpha_{EW}\mu\zeta_i}\over{16{\sqrt2}{\pi}s_W^2c_W^2s_\beta\Lambda^2}}(c_W(U_{\chi^0})_{2k}-s_W(U_{\chi^0})_{1k})m_{\chi_k^0}$.
 Additionally, the radiative corrections to the mixing between left- and right-handed neutrinos are proportional to $Y_N\upsilon_{L_i}$ or $A_N\upsilon_{L_i}$ and can be ignored safely.

Considering those one-loop corrections, the mass matrix in Eq. (15) is rewritten  as
\begin{eqnarray}
&&M'=\left(\begin{array}{ccc}
(\Delta{\cal M}_L^{(0)})_{3\times3} & ({\cal A}_N^{(1)}U_N)_{3\times4} & ({\cal A}_N^{(2)})_{3\times4}+\Delta{\cal A}_N^{(2)}\\
(U_N^T{\cal A}_N^{(1)T})_{4\times3} & (U_N^T{\cal M}_N^{(0)}U_N)_{4\times4}+(\Delta{\cal M}_N^{(0)})_{4\times4} & (U_N^T{\cal A}_N^{(3)})_{4\times4}\\
({\cal A}_N^{(2)T})_{4\times3} & ({\cal A}_N^{(3)T}U_N)_{4\times4} & ({\cal M}_N)_{4\times4}
\end{array}\right)\nonumber\\
&&\hspace{0.8cm}=\left(\begin{array}{cc}
(m_\nu+\Delta(m_\nu))_{5\times5} & (m_D+\Delta(m_D))_{5\times6}\\
(m_D+\Delta(m_D))^T_{6\times5} & ({\cal M}+\Delta({\cal M}))_{6\times6}
\end{array}\right).
\label{UNTMUN}
\end{eqnarray}
Using the seesaw mechanism, the effective mass matrix for five light neutrinos (three active and two sterile) at the one-loop level is
\begin{eqnarray}
&&{m'_{eff}}\simeq{(m_\nu+\Delta(m_\nu))}-(m_D+\Delta(m_D))\cdot{({\cal M}+\Delta({\cal M}))}^{-1}\cdot{(m_D+\Delta(m_D))^T}\nonumber\\
&&\hspace{1.6cm}-{1\over2}(m_\nu+\Delta(m_\nu))\cdot{(m_D+\Delta(m_D))\cdot{({\cal M}+\Delta({\cal M}))}^{-2}}\cdot{(m_D+\Delta(m_D))^T}\nonumber\\
&&\hspace{1.6cm}-{1\over2}(m_D+\Delta(m_D))\cdot{({\cal M}+\Delta({\cal M}))}^{-2}\cdot{(m_D+\Delta(m_D))^T}\cdot{(m_\nu+\Delta(m_\nu))}\nonumber\\
&&\hspace{1.4cm}\simeq\left(\begin{array}{cc}
[{M'}_\nu^{LL}]_{3\times3} & [{M'}_\nu^{LR}]_{3\times2}\\
{[{M'}_\nu^{LR}]^T_{2\times3}} & [{M'}_\nu^{RR}]_{2\times2}
\end{array}\right).
\label{meff}
\end{eqnarray}

 We can obtain five eigenvalues by diagonalizing the effective mass matrix $m'_{eff}$. The corrections to the sterile neutrinos are much larger than the corrections to the active neutrinos which are suppressed by the tiny parameters, so the three light eigenvalues are active neutrinos, and other two relatively heavy eigenvalues are sterile neutrinos. Under the assumption $[{M'}_\nu^{RR}]\gg[{M'}_\nu^{LL}]$, the corrected effective mass matrix of three left-handed neutrinos is
\begin{eqnarray}
{m'}_{\nu_L}^{eff}\simeq{{M'}_\nu^{LL}}.
\label{H}
\end{eqnarray}
Using the "top-down" method \cite{top-down,zhang} in the effective mass matrix ${m'}_{\nu_L}^{eff}$, we diagonalize the Hermitian matrix
\begin{eqnarray}
{\cal{H}}=({m'}_{\nu_L}^{eff})^\dag{{m'}_{\nu_L}^{eff}}.
\label{H}
\end{eqnarray}
The eigenvalues of the $3\times3$ effective mass squared matrix $\cal{H}$ are given as
\begin{eqnarray}
&&m_1^2={a\over3}-{1\over3}p(\cos\phi+{\sqrt3}\sin\phi),\nonumber\\
&&m_2^2={a\over3}-{1\over3}p(\cos\phi-{\sqrt3}\sin\phi),\nonumber\\
&&m_3^2={a\over3}+{2\over3}p\cos\phi.
\label{eigenvalues}
\end{eqnarray}
To formulate the expressions of a concise form, one can define the notations
\begin{eqnarray}
&&p=\sqrt{a^2-3b},\quad \phi={1\over3}\arccos({1\over{p^3}}(a^3-{9\over2}ab+{27\over2}c)),\qquad a=\mathrm{Tr}(\cal{H}),\nonumber\\
&&b={\cal{H}}_{11}{\cal{H}}_{22}+{\cal{H}}_{11}{\cal{H}}_{33}+{\cal{H}}_{22}{\cal{H}}_{33}-{\cal{H}}_{12}^2-{\cal{H}}_{13}^2-{\cal{H}}_{23}^2,
\quad c=\mathrm{Det}(\cal{H}).
\label{pabc}
\end{eqnarray}
For the three active neutrino mixing, there are two possible solutions on the neutrino mass spectrum. The normal ordering (NO) spectrum is
\begin{eqnarray}
&&m_{\nu_1}<m_{\nu_2}<m_{\nu_3},\quad m_{\nu_1}^2=m_1^2,\quad m_{\nu_2}^2=m_2^2,\quad m_{\nu_3}^2=m_3^2,\nonumber\\
&&\Delta{m_\odot^2}=m_{\nu_2}^2-m_{\nu_1}^2,\quad \Delta{m_A^2}=m_{\nu_3}^2-m_{\nu_1}^2,
\label{NO}
\end{eqnarray}
and the neutrino mass spectrum with the inverted ordering (IO) is
\begin{eqnarray}
&&m_{\nu_3}<m_{\nu_1}<m_{\nu_2},\quad m_{\nu_3}^2=m_1^2,\quad m_{\nu_1}^2=m_2^2,\quad m_{\nu_2}^2=m_3^2,\nonumber\\
&&\Delta{m_\odot^2}=m_{\nu_2}^2-m_{\nu_1}^2,\quad \Delta{m_A^2}=m_{\nu_2}^2-m_{\nu_3}^2.
\label{IO}
\end{eqnarray}
From the mass squared matrix $\cal{H}$ and three eigenvalues one can get the orthogonal matrix $U_\nu$ of $\cal{H}$ \cite{top-down,zhang}. Correspondingly, the mixing angles among three active neutrinos are determined by
\begin{eqnarray}
&&\sin\theta_{13}=|(U_\nu)_{13}|,\qquad \sin\theta_{23}={{|(U_\nu)_{23}|}\over{\sqrt{1-|(U_\nu)_{13}|^2}}},
\qquad\sin\theta_{12}={{|(U_\nu)_{12}|}\over{\sqrt{1-|(U_\nu)_{13}|^2}}}.
\label{angles}
\end{eqnarray}

It is important to calculate the active-sterile neutrino mixing angles which are strongly constrained by X-ray observations \cite{DM}. There are several mixing angles $\theta_{\sigma I}$, where $I$ is the sterile neutrino flavor and $\sigma$ is the active neutrino flavor. We define $\theta_{\sigma I}^2=({M'}_\nu^{LR})_{\sigma I}^2/m_{rI}^2$, where $m_{rI}^2$ are the sterile neutrino masses. There are two sterile neutrinos; then, $I=1,2$. We define the active-sterile neutrino mixing angle as \cite{abundance}
\begin{eqnarray}
\theta_I^2=\sum_{\sigma=e,\mu,\tau}\theta_{\sigma I}^2.
\label{angles}
\end{eqnarray}

\section{Numerical results\label{sec5}}

The neutrino oscillation experimental data \cite{neutrino-number} and the lightest \textit{CP}-even Higgs $h^0$ with a mass $m_{h^0}\simeq125\;\rm{GeV}$ \cite{Higgs} constrain relevant parameter space strongly. In numerical analysis, we adopt the relevant parameters as default,
\begin{eqnarray}
&&\upsilon_{N_3}=2\;{\rm{TeV}}, \qquad M_1=1.2\;{\rm{TeV}},\qquad M_2=1.6\;{\rm{TeV}},\qquad m_{BBL}=1.3\;{\rm{TeV}}, \nonumber\\
&& m_{A}= 1 \;{\rm{TeV}},\qquad g_{BL}=0.6, \qquad\quad \mu=-1\;{\rm{TeV}}, \quad (m_{\tilde{N}^c}^2)_{11}=(m_{\tilde{N}^c}^2)_{22}=1.8\;{\rm{TeV}},\nonumber\\
&&m_{\tilde{\nu}_{L_1}}=2100\;{\rm{GeV}},\qquad m_{\tilde{\nu}_{L_2}}=2200\;{\rm{GeV}},\qquad  m_{\tilde{\nu}_{L_3}}=2300\;{\rm{GeV}},\nonumber\\
&&m_{\tilde{e}_{L_1}}=3100\;{\rm{GeV}},\qquad m_{\tilde{e}_{L_2}}=3200\;{\rm{GeV}},\qquad  m_{\tilde{e}_{L_3}}=3300\;{\rm{GeV}}
\label{parameters}
\end{eqnarray}
to reduce the number of free parameters in the model considered here.

As mentioned above, the masses of two light sterile neutrinos mainly originate from one-loop radiative corrections. The corrections to the sterile neutrino masses depend on two heavy Majorana fermion masses $m_{N_\alpha}$, the orthogonal matrix $U_N$ and $U(1)_{B-L}$ gauge boson mass $m_{Z_{BL}}$ from Eq. (29). The mass $m_{Z_{BL}}$ depends on the VEVs $\upsilon_{N_i}$ of right-handed sneutrinos. From Eqs. (13) and (14), both $m_{N_\alpha}$ and $U_N$ are determined by $\upsilon_{N_i}$ and $m_{BL}$ which is the $U(1)_{B-L}$ gaugino mass in soft breaking terms.  At the same time, these parameters affect the active neutrino masses from Eq. (34). In this section, we analyze the numerical results for the mixing angles and mass squared differences of active neutrinos varying with $\upsilon_{N_2}$, $m_{BL}$, and $\tan\beta$, assuming neutrino mass spectrum with normal ordering (NO) and inverted ordering (IO). Meanwhile, we discuss the numerical results of two sterile neutrino masses and the active-sterile neutrino mixing angles varying with these parameters.

\subsection{NO spectrum}
In order to fit the experimental data on active neutrino mass squared differences and mixing angles in this scenario, we choose the VEVs of left-handed sneutrinos and the Yukawa couplings of right-handed neutrinos, respectively, as
\begin{eqnarray}
&&\upsilon_{L_1}=1.547\times10^{-4}\;{\rm{GeV}}, \qquad  Y_1=9.982\times10^{-7}, \nonumber\\
&&\upsilon_{L_2}=3.220\times10^{-4}\;{\rm{GeV}}, \qquad  Y_2=2.440\times10^{-6}, \nonumber\\
&&\upsilon_{L_3}=1.256\times10^{-4}\;{\rm{GeV}}, \qquad  Y_1=1.468\times10^{-6}.
\label{parameters}
\end{eqnarray}
 Correspondingly the theoretical predictions on active neutrino mixing angles, mass squared differences, the sum of the active neutrino masses, two sterile neutrino masses $m_{r1}$, $m_{r2}$, and the active-sterile neutrino mixing angles $\theta_1^2$, $\theta_2^2$ are derived as
\begin{eqnarray}
&&\sin^2\theta_{12}=0.3072, \qquad\qquad\quad  \sin^2\theta_{23}=0.4370, \qquad\qquad\quad   \sin^2\theta_{13}=0.0234,\nonumber\\
&&\Delta m_A^2=2.430\times10^{-3}\;{\rm{eV^2}}, \quad \Delta m_\odot^2=7.532\times10^{-5}\;{\rm{eV^2}},\quad \sum\limits_i m_{\nu_i}=5.798\times10^{-2}\;{\rm{eV}},  \nonumber\\
&&m_{r1}=7.13\;{\rm{KeV}},\qquad\qquad\quad m_{r2}=12.88\;{\rm{KeV}}, \nonumber\\
&&\theta_1^2=2.58\times10^{-11},\qquad\qquad\quad \theta_2^2=1.17\times10^{-10},
\label{parameters}
\end{eqnarray}
when $\upsilon_{N_1}=3\;\rm{GeV}$, $\upsilon_{N_2}=7.7\;\rm{GeV}$, $m_{BL}=1.08\;\rm{TeV}$, and $\tan\beta=20$.

%%%%%%%%%%%%%%%%%%%%%%%%%%%%%%%%%%%%%%%%%%%%%%%%%%%%%
\begin{figure}[h]
\setlength{\unitlength}{1mm}
\centering
\includegraphics[width=2.5in]{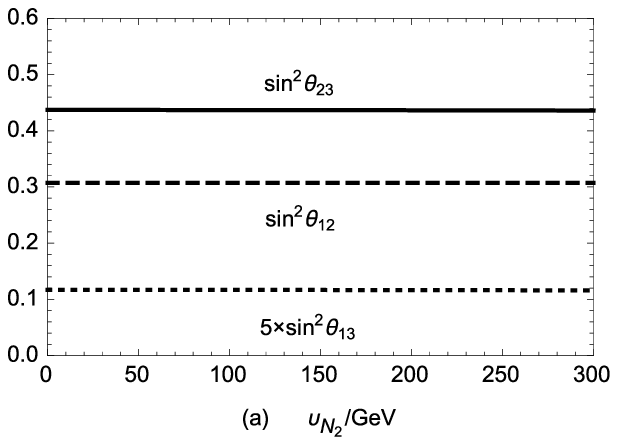}
\vspace{0.0cm}
\includegraphics[width=2.5in]{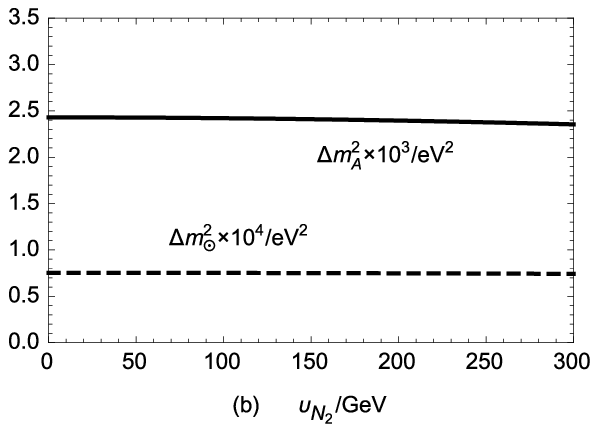}
\vspace{0.0cm}
\includegraphics[width=2.5in]{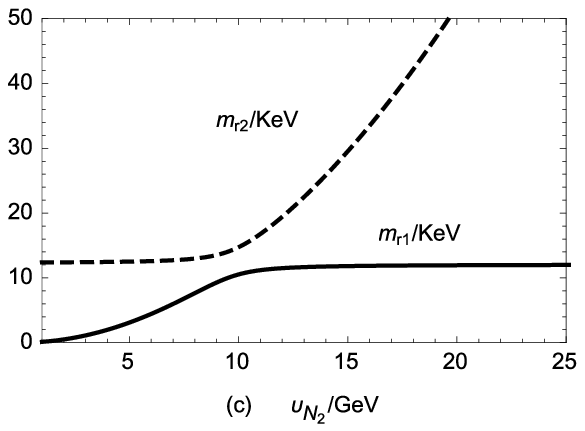}
\vspace{0.0cm}
\includegraphics[width=2.5in]{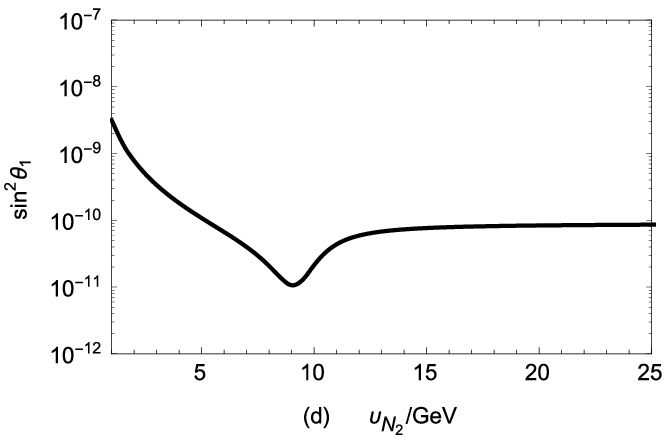}
\vspace{0cm}
\caption[]{Assuming neutrino mass spectrum with NO, we plot the mixing angles, mass squared differences of active neutrinos, two sterile neutrino masses, and active-sterile neutrino mixing angle versus the VEV $\upsilon_{N_2}$ of right-handed sneutrinos, where (a) the solid line stands for $\sin^2{\theta_{23}}$ versus $\upsilon_{N_2}$, the dashed line for stands $\sin^2{\theta_{12}}$ versus $\upsilon_{N_2}$, and the dotted line stands for $\sin^2{\theta_{13}}$ versus $\upsilon_{N_2}$, (b) the solid line stands for $\Delta{m_A^2}$ versus $\upsilon_{N_2}$ and the dashed line stands for $\Delta{m_{\odot}^2}$ versus $\upsilon_{N_2}$, (c) the solid line stands for $m_{r1}$ versus $\upsilon_{N_2}$ and the dashed line stands for $m_{r2}$ versus $\upsilon_{N_2}$, and (d) the solid line stands for $\sin^2\theta_1$ of active-sterile neutrino mixing angle  versus $\upsilon_{N_2}$.}
\label{fig1}
\end{figure}
%%%%%%%%%%%%%%%%%%%%%%%%%%%%%%%%%%%%%%%%%%%%%%%%%%%%%
Assuming neutrino mass spectrum with NO and taking $\upsilon_{N_1}=3\;\rm{GeV}$, $m_{BL}=1.08\;\rm{TeV}$, and $\tan\beta=20$, we depict the active neutrino mixing angles varying with the VEV $\upsilon_{N_2}$ of right-handed sneutrinos in Fig. 1(a), where the solid line denotes  $\sin^2{\theta_{23}}$ versus $\upsilon_{N_2}$, the dashed line denotes $\sin^2{\theta_{12}}$ versus $\upsilon_{N_2}$, and the dotted line denotes $\sin^2{\theta_{13}}$ versus $\upsilon_{N_2}$. With the increasing of $\upsilon_{N_2}$, theoretical predictions of these mixing angles vary gently. In this region of $\upsilon_{N_2}$, the three mixing angles satisfy the experiment bounds simultaneously \cite{neutrino-number}. Using the same choice on parameter space, we draw the mass squared differences of active neutrinos varying with $\upsilon_{N_2}$ in Fig. 1(b), where the solid line denotes $\Delta{m_A^2}$ versus $\upsilon_{N_2}$, and the dashed line denotes $\Delta{m_{\odot}^2}$ versus $\upsilon_{N_2}$. With the increasing of $\upsilon_{N_2}$, $\Delta{m_A^2}$ and $\Delta{m_{\odot}^2}$ decreases slowly. The effective mass matrix for active neutrinos depends on $\upsilon_{N_2}$ through the term   $\zeta_i\zeta_j/\Lambda_\zeta\simeq(2\tilde{\mu}^4\upsilon_u^2{m_{_{BL}}})/(\Lambda_{\tilde{m}^4}\upsilon_N^2)$, and $\upsilon_{N_2}$ has relatively small influence on $\upsilon_{N}$ ($\upsilon_{N_2}\ll\upsilon_{N}$).

 Additionally, we study the two sterile neutrino masses $m_{r1}$, $m_{r2}$ varying with $\upsilon_{N_2}$ in Fig. 1(c), where the solid line denotes $m_{r1}$ versus $\upsilon_{N_2}$, and the dashed line denotes $m_{r2}$ versus $\upsilon_{N_2}$. It shows that two sterile neutrinos obtain $\rm{KeV}$ scale masses. When $\upsilon_{N_2}\leq{10\;\rm{GeV}}$, $m_{r1}$ increases steeply with the increasing of $\upsilon_{N_2}$, and $m_{r2}$ changes mildly with $\upsilon_{N_2}$. However, when $\upsilon_{N_2}\geq{10\;\rm{GeV}}$, the dependence of $m_{r1}$ on $\upsilon_{N_2}$ is not obvious, and $m_{r2}$ increases quickly with the increasing of $\upsilon_{N_2}$. This is because the two sterile neutrinos obtain masses from the one-loop corrections. The corrections depend on $U_N$, $m_{N_\alpha}$, and $m_{Z_{BL}}^2$, which are all related to $\upsilon_{N_2}$. Under the same choice on parameter space, the numerical result of the active-heavier sterile neutrino mixing angle $\sin^2\theta_2$ changes gently about $10^{-10}$ or $10^{-11}$. We only study the active-lighter sterile neutrino mixing angle $\sin^2\theta_1$ varying with $\upsilon_{N_2}$ in Fig. 1(d). Considering the restrictions of X-ray line searches on the mixing angle, the applicable range of $\upsilon_{N_2}$ is about from $5$ to ${12\;\rm{GeV}}$  \cite{DM}. When $\upsilon_{N_2}=7.7\;\rm{GeV}$, the sterile neutrino mass $m_{r1}$ is about $7.13\;{\rm{KeV}}$ with the mixing angle $\sin^2\theta_1\sim10^{-11}$ which can explain the observed X-ray line at $3.5\;{\rm{KeV}}$ \cite{sterile neutrino1,sterile neutrino2}. So, the lighter sterile neutrino can be a dark matter candidate.

\begin{figure}[h]
\setlength{\unitlength}{1mm}
\centering
\includegraphics[width=2.5in]{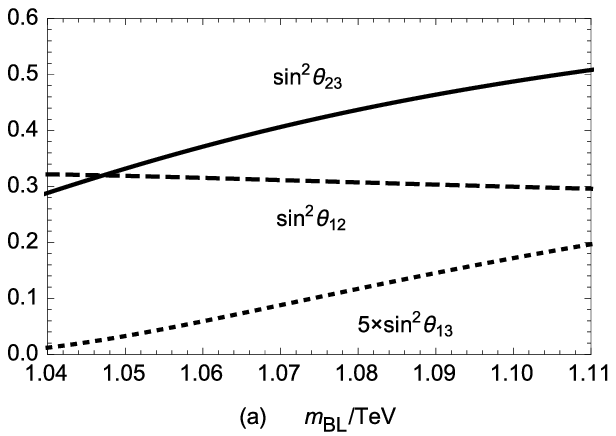}
\vspace{0.0cm}
\includegraphics[width=2.5in]{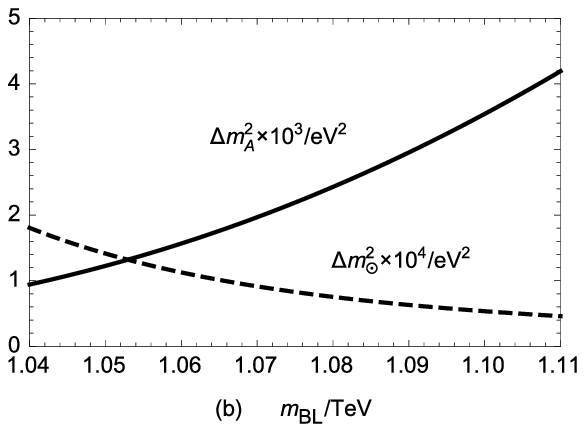}
\vspace{0cm}
\includegraphics[width=2.5in]{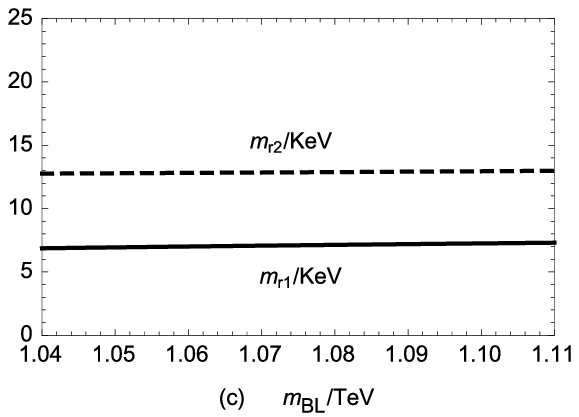}
\vspace{0cm}
\includegraphics[width=2.5in]{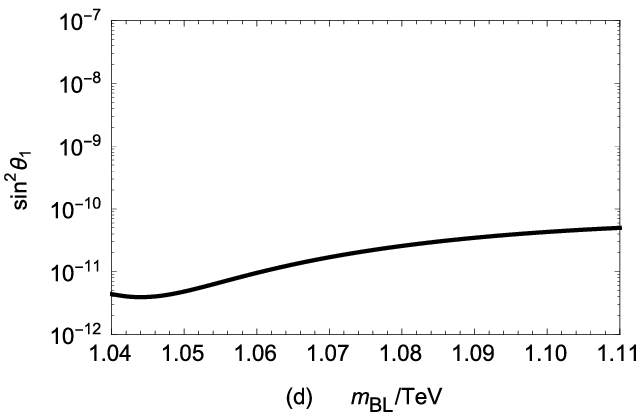}
\vspace{0cm}
\caption[]{Assuming neutrino mass spectrum with NO, we plot the mixing angles, mass squared differences of active neutrinos, two sterile neutrino masses and active-sterile neutrino mixing angle versus $U(1)_{B-L}$ gaugino mass $m_{BL}$, where (a) the solid line stands for $\sin^2{\theta_{23}}$ versus $m_{BL}$, the dashed line stands for $\sin^2{\theta_{12}}$ versus $m_{BL}$, and the dotted line stands for $\sin^2{\theta_{13}}$ versus $m_{BL}$, (b) the solid line stands for $\Delta{m_A^2}$ versus $m_{BL}$ and the dashed line stands for $\Delta{m_{\odot}^2}$ versus $m_{BL}$, (c) the solid line stands for $m_{r1}$ versus $m_{BL}$ and the dashed line stands for $m_{r2}$ versus $m_{BL}$, and (d) the solid line stands for $\sin^2\theta_1$ versus $m_{BL}$.}
\label{fig2}
\end{figure}
In this model, the $U(1)_{B-L}$ gaugino mass $m_{BL}$ also affects the final numerical results of the neutrino sector. Taking $\tan\beta=20$ and $\upsilon_{N_1}=3\;\rm{GeV}$, $\upsilon_{N_2}=7.7\;\rm{GeV}$; we plot the active neutrino mixing angles varying with $m_{BL}$ in Fig. 2(a), where the solid line denotes $\sin^2{\theta_{23}}$ versus $\upsilon_{N_2}$, the dashed line denotes $\sin^2{\theta_{12}}$ versus $\upsilon_{N_2}$, and the dotted line denotes $\sin^2{\theta_{13}}$ versus $\upsilon_{N_2}$. With the increasing of $m_{BL}$, the theoretical prediction on the mixing angle $\sin^2{\theta_{12}}$ depends on $m_{BL}$ mildly, and the mixing angles $\sin^2{\theta_{23}}$ , $\sin^2{\theta_{13}}$ increase steeply. Using the same choice on parameter space, we plot the mass squared differences of active neutrinos varying with $m_{BL}$ in Fig. 2(b), where the solid line denotes $\Delta{m_A^2}$ versus $\upsilon_{N_2}$, and the dashed line denotes $\Delta{m_{\odot}^2}$ versus $\upsilon_{N_2}$.  It shows that $\Delta{m_A^2}$ raises steeply with the increasing of $m_{BL}$, and $\Delta{m_{\odot}^2}$ diminishes quickly with the increasing of $m_{BL}$. The effective mass matrix for active neutrinos depends on $m_{BL}$ through the term $\zeta_i\zeta_j/\Lambda_\zeta\simeq(2\tilde{\mu}^4\upsilon_u^2{m_{_{BL}}})/(\Lambda_{\tilde{m}^4}\upsilon_N^2)$; therefore, the numerical evaluations on $\sin^2{\theta_{23}}$, $\sin^2{\theta_{13}}$, $\Delta{m_A^2}$ and $\Delta{m_{\odot}^2}$ depend on $m_{BL}$ strongly. From those numerical results on these parameter spaces, we find that the updated experiment data require $m_{BL}\sim1.08\;\rm{TeV}$. Additionally we study the masses of two sterile neutrinos varying with  $m_{BL}$ in Fig. 2(c). The numerical results indicate that both $m_{r1}$ and $m_{r2}$ depend on $m_{BL}$ mildly. The active-sterile neutrino mixing angle $\sin^2\theta_2$ changes gently about $10^{-10}$ with the increasing of $m_{BL}$. We study the active-sterile neutrino mixing angle $\sin^2\theta_1$ varying with $\upsilon_{N_2}$ in Fig. 2(d). With increasing of $m_{BL}$, the mixing angle $\sin^2\theta_1$ increases quickly.

\begin{figure}[h]
\setlength{\unitlength}{1mm}
\centering
\includegraphics[width=2.5in]{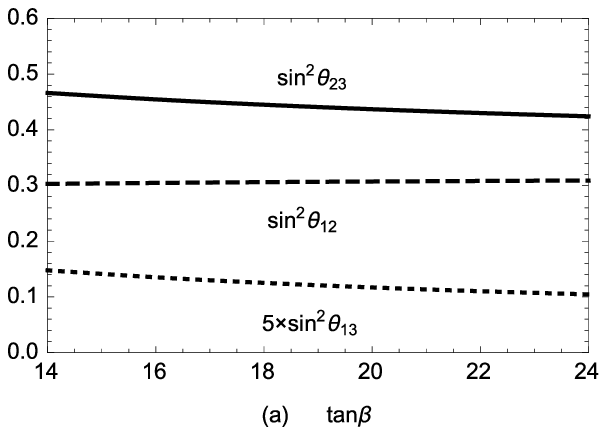}
\vspace{0.0cm}
\includegraphics[width=2.5in]{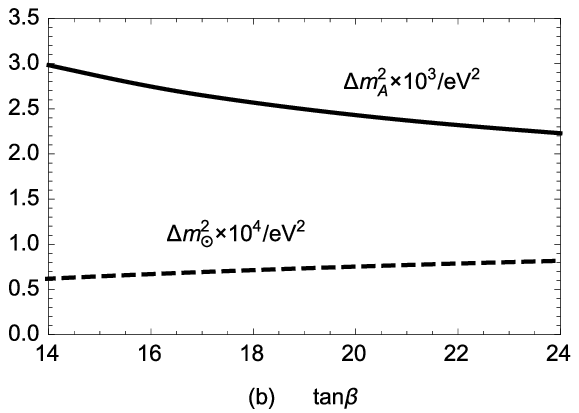}
\vspace{0.0cm}
\includegraphics[width=2.5in]{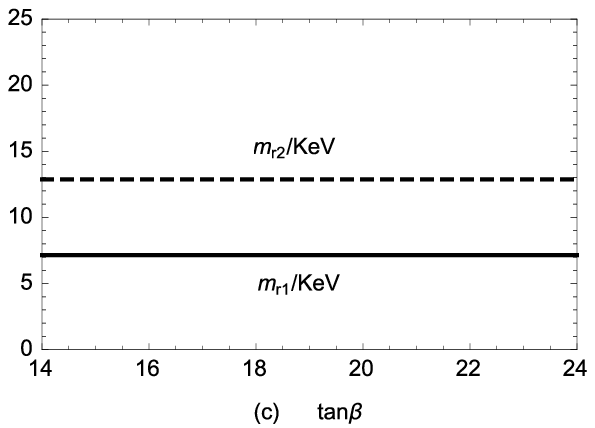}
\vspace{0cm}
\includegraphics[width=2.5in]{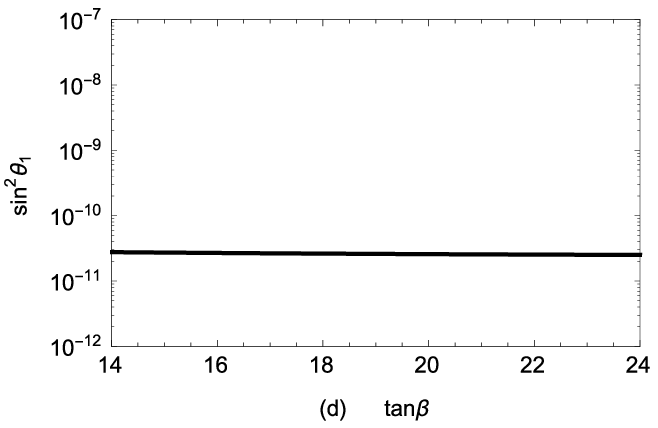}
\vspace{0cm}
\caption[]{Assuming neutrino mass spectrum with NO, we plot the mixing angles, mass squared differences of active neutrinos, two sterile neutrino masses and active-sterile neutrino mixing angle versus $\tan\beta$, where (a) the solid line stands for $\sin^2{\theta_{23}}$ versus $\tan\beta$, the dashed line stands for $\sin^2{\theta_{12}}$ versus $\tan\beta$, and the dotted line stands for $\sin^2{\theta_{13}}$ versus $\tan\beta$, (b) the solid line stands for $\Delta{m_A^2}$ versus $\tan\beta$ and the dashed line stands for $\Delta{m_{\odot}^2}$ versus $\tan\beta$, (c) the solid line stands for $m_{r1}$ versus $\tan\beta$ and the dashed line stands for $m_{r2}$ versus $\tan\beta$, and (d) the solid line stands for $\sin^2\theta_1$ versus $\tan\beta$.}
\label{fig3}
\end{figure}
Taking $m_{BL}=1.08\;\rm{TeV}$, $\upsilon_{N_1}=3\;\rm{GeV}$, and $\upsilon_{N_2}=7.7\;\rm{GeV}$, we draw the active neutrino mixing angles varying with  $\tan\beta$ in Fig. 3(a), where the solid line denotes $\Delta{m_A^2}$ versus $\upsilon_{N_2}$, and the dashed line denotes $\Delta{m_{\odot}^2}$ versus $\upsilon_{N_2}$. With the increasing of $\tan\beta$, theoretical predictions on the mixing angle $\sin^2{\theta_{12}}$ varies gently, and the mixing angles $\sin^2{\theta_{23}}$ , and $\sin^2{\theta_{13}}$ decrease slowly. Using the same choice on parameter space, we draw the mass squared differences of active neutrinos varying with $\tan\beta$ in Fig. 3(b), where the solid line denotes $\Delta{m_A^2}$ versus $\upsilon_{N_2}$, and the dashed line denotes $\Delta{m_{\odot}^2}$ versus $\upsilon_{N_2}$. It shows that $\Delta{m_{\odot}^2}$  varies mildly and $\Delta{m_A^2}$ decreases steeply with the increasing of $\tan\beta$. Additionally, we study the masses of two sterile neutrinos versus $\tan\beta$ in Fig. 3(c); the numerical result implies that two sterile neutrino masses depend on $\tan\beta$ gently. This is because two sterile neutrinos obtain masses mainly from the radiative corrections, and from Eq. (29) the corrections on masses of two sterile neutrinos are almost not dependent on $\tan\beta$. For the same reason, the active-sterile neutrino mixing angle has barely changed with $\tan\beta$. The numerical result of $\sin^2\theta_2$ is about $10^{-10}$. The numerical result of $\sin^2\theta_1$ is about $10^{-11}$ in Fig. 3(d).

Assuming neutrino mass spectrum with normal ordering, two sterile neutrinos obtain $\rm{KeV}$ scale masses. Both $\rm{KeV}$ sterile neutrinos were produced in the early Universe via oscillations. The lighter sterile neutrino forms dark matter, but the oscillation mechanism cannot produce enough of these neutrinos to act as all dark matter for given $m_{r1}$ and $\sin^2\theta_1$ \cite{neutrino DM,ShF}. The heavier sterile neutrino can decay into the lighter one to enrich the production of the sterile neutrino DM.

\subsection{IO spectrum}
With the active neutrino mass spectrum being IO spectrum, we choose the VEVs of left-handed sneutrinos and the Yukawa couplings of sterile neutrinos, respectively, as
\begin{eqnarray}
&&\upsilon_{L_1}=1.899\times10^{-4}\;{\rm{GeV}}, \qquad  Y_1=4.989\times10^{-7}, \nonumber\\
&&\upsilon_{L_2}=4.208\times10^{-4}\;{\rm{GeV}}, \qquad  Y_2=2.896\times10^{-6}, \nonumber\\
&&\upsilon_{L_3}=3.434\times10^{-4}\;{\rm{GeV}}, \qquad  Y_3=2.551\times10^{-6}.
\label{parameters}
\end{eqnarray}
Correspondingly the theoretical predictions on active neutrino mixing angles, mass squared differences, the sum of the active neutrino masses, two sterile neutrino masses $m_{r1}$, $m_{r2}$, and the active-sterile neutrino mixing angles $\theta_1^2$, $\theta_2^2$ are derived as
\begin{eqnarray}
&&\sin^2\theta_{12}=0.3077, \qquad\qquad\quad  \sin^2\theta_{23}=0.4556, \qquad\qquad\quad  \sin^2\theta_{13}=0.0243,\nonumber\\
&&\Delta m_A^2=2.381\times10^{-3}\;{\rm{eV^2}}, \quad \Delta m_\odot^2=7.627\times10^{-5}\;{\rm{eV^2}},\quad \sum\limits_i m_{\nu_i}=9.682\times10^{-2}\;{\rm{eV}},  \nonumber\\
&&m_{r1}=7.13\;{\rm{KeV}},\qquad\qquad\quad m_{r2}=12.88\;{\rm{KeV}},\nonumber\\
&&\theta_1^2=2.84\times10^{-11},\qquad\qquad\quad \theta_2^2=3.27\times10^{-10},
\label{parameters}
\end{eqnarray}
when $\upsilon_{N_1}=3\;\rm{GeV}$, $\upsilon_{N_2}=7.7\;\rm{GeV}$, $m_{BL}=1.08\;\rm{TeV}$, and $\tan\beta=20$.
%%%%%%%%%%%%%%%%%%%%%%%%%%%%%%%%%%%%%%%%%%%%%%%%%%%%%
\begin{figure}[h]
\setlength{\unitlength}{1mm}
\centering
\includegraphics[width=2.5in]{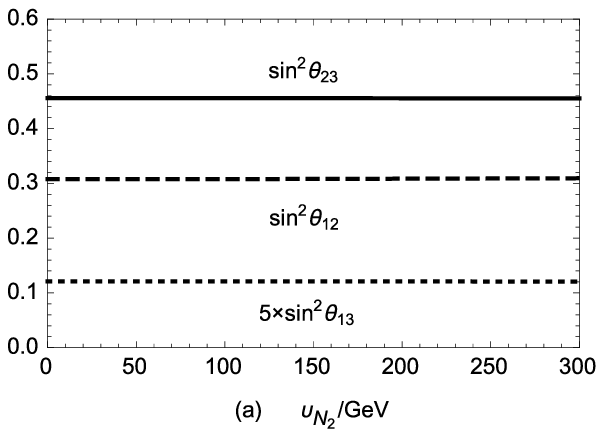}
\vspace{0.0cm}
\includegraphics[width=2.5in]{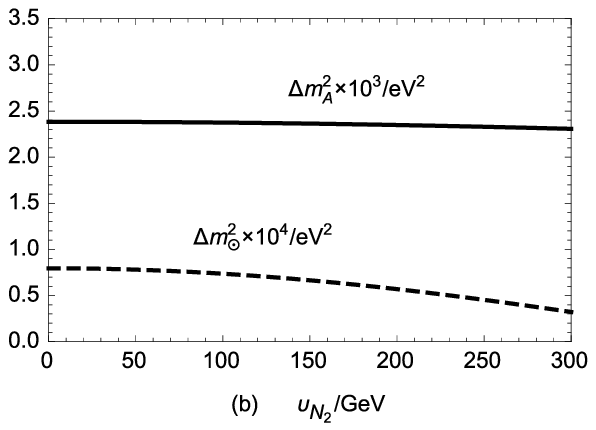}
\vspace{0cm}
\includegraphics[width=2.5in]{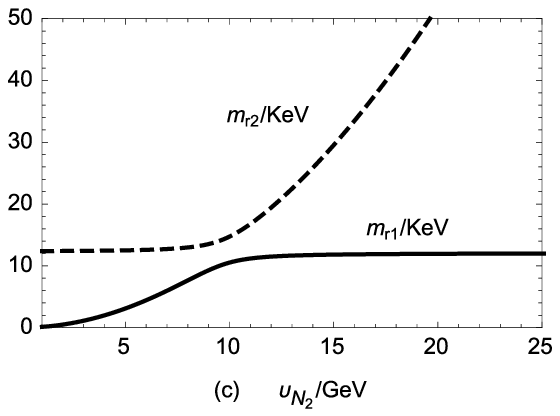}
\vspace{0cm}
\includegraphics[width=2.5in]{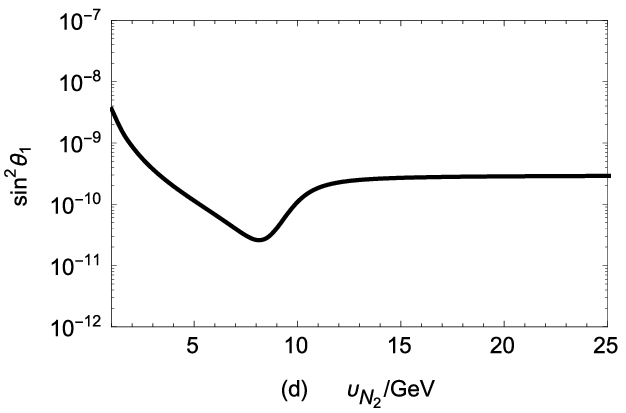}
\vspace{0cm}
\caption[]{Assuming neutrino mass spectrum with IO, we plot the mixing angles, mass squared differences of active neutrinos, two sterile neutrino masses and active-sterile neutrino mixing angle versus the VEV $\upsilon_{N_2}$ of right-handed sneutrinos, where (a) the solid line stands for $\sin^2{\theta_{23}}$ versus $\upsilon_{N_2}$, the dashed line stands for $\sin^2{\theta_{12}}$ versus $\upsilon_{N_2}$, and the dotted line stands for $\sin^2{\theta_{13}}$ versus $\upsilon_{N_2}$, (b) the solid line stands for $\Delta{m_A^2}$ versus $\upsilon_{N_2}$ and the dashed line stands for $\Delta{m_{\odot}^2}$ versus $\upsilon_{N_2}$, (c) the solid line stands for $m_{r1}$ versus $\upsilon_{N_2}$ and the dashed line stands for $m_{r2}$ versus $\upsilon_{N_2}$, and (d) the solid line stands for $\sin^2\theta_1$ of active-sterile neutrino mixing angle versus $\upsilon_{N_2}$.}
\label{fig4}
\end{figure}
%%%%%%%%%%%%%%%%%%%%%%%%%%%%%%%%%%%%%%%%%%%%%%%%%%%%%

When the neutrino mass spectrum is IO, the manners of parameters $\upsilon_{N_2}$, $m_{BL}$, and $\tan\beta$ affecting the numerical results on the neutrino sector may differ from that of the neutrino mass spectrum with NO. Assuming neutrino mass spectrum with IO and taking $\upsilon_{N_1}=3\;\rm{GeV}$, $m_{BL}=1.08\;\rm{TeV}$, and $\tan\beta=20$, we depict the active neutrino mixing angles varying with $\upsilon_{N_2}$ in Fig. 4(a), where the solid line denotes  $\sin^2{\theta_{23}}$ versus $\upsilon_{N_2}$, the dashed line denotes $\sin^2{\theta_{12}}$ versus $\upsilon_{N_2}$, and the dotted line denotes $\sin^2{\theta_{13}}$ versus $\upsilon_{N_2}$. Obviously, theoretical predictions on those mixing angles vary slowly with the increasing of $\upsilon_{N_2}$. Adopting the same choice on parameter space, we draw the mass squared differences of active neutrinos varying with  $\upsilon_{N_2}$ in Fig. 4(b), where the solid line denotes  $\Delta{m_A^2}$  versus $\upsilon_{N_2}$, and the dashed line denotes $\Delta{m_{\odot}^2}$ versus $\upsilon_{N_2}$. It shows that $\Delta{m_A^2}$ decreases gently with the increasing of $\upsilon_{N_2}$, but $\Delta{m_{\odot}^2}$ decreases steeply. Taking into account the neutrino experiment bounds, the appropriate region of $\upsilon_{N_2}$ is $\upsilon_{N_2}\leq60\;\rm{GeV}$. In addition, we study the masses of two sterile neutrinos varying with $\upsilon_{N_2}$ in Fig. 4(c), where the solid line denotes $m_{r1}$ versus $\upsilon_{N_2}$, and the dashed line denotes $m_{r2}$ versus $\upsilon_{N_2}$. It shows that two sterile neutrinos masses have almost the same changing trend as the NO spectrum. Because two sterile neutrinos obtain relatively large masses than active neutrinos, they are almost irrelevant to the active neutrino mass spectrum. Under the same choice on parameter space, the numerical result of the active-heavier sterile neutrino mixing angle $\sin^2\theta_2$ changes gently about $10^{-10}$. We only study the active-lighter sterile neutrino mixing angle $\sin^2\theta_1$ versus $\upsilon_{N_2}$ in Fig. 4(d). It shows that the mixing angle $\sin^2\theta_1$ depends on $\upsilon_{N_2}$ strongly. Considering the restrictions of X-ray line searches on the mixing angle, the applicable range of $\upsilon_{N_2}$ is about from $5$ to ${10\;\rm{GeV}}$ \cite{DM}. When $\upsilon_{N_2}=7.7\;\rm{GeV}$, the sterile neutrino mass $m_{r1}$ is about $7.13\;{\rm{KeV}}$ with the mixing angle $\sin^2\theta_1\sim10^{-11}$ which can explain the observed X-ray line at $3.5\;{\rm{KeV}}$ \cite{sterile neutrino1,sterile neutrino2}.
Therefore, the lighter sterile neutrino can be a dark matter candidate.

\begin{figure}[h]
\setlength{\unitlength}{1mm}
\centering
\includegraphics[width=2.5in]{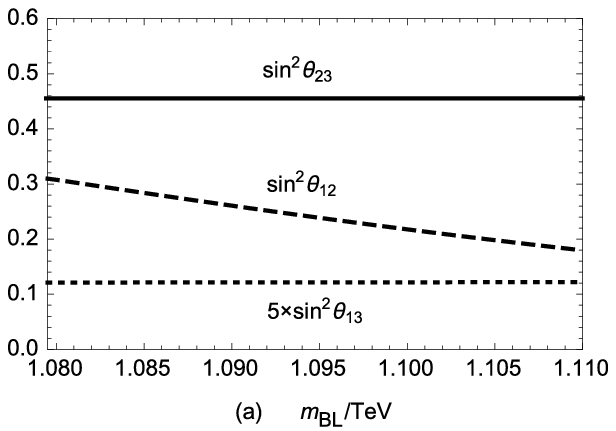}
\vspace{0.0cm}
\includegraphics[width=2.5in]{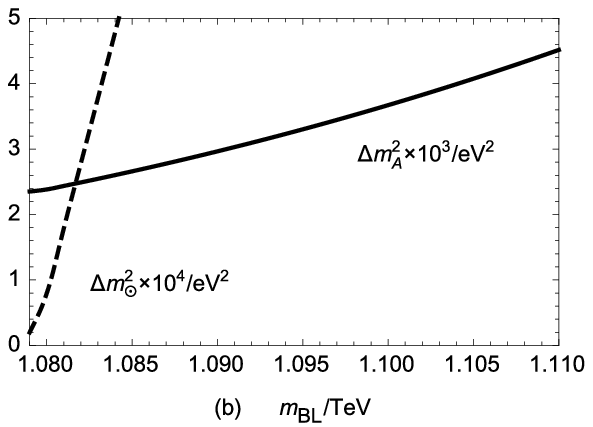}
\vspace{0cm}
\includegraphics[width=2.5in]{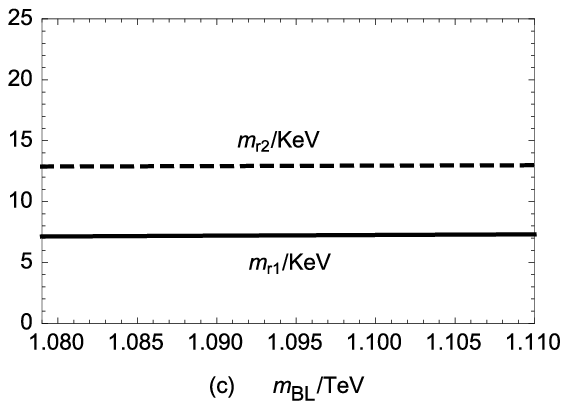}
\vspace{0cm}
\includegraphics[width=2.5in]{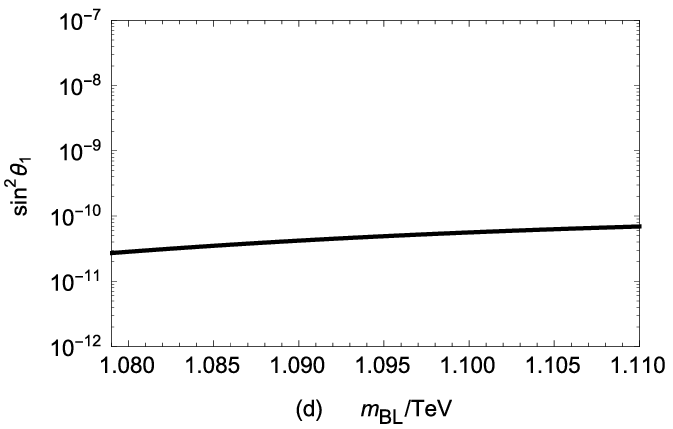}
\vspace{0cm}
\caption[]{Assuming neutrino mass spectrum with IO, we plot the mixing angles, mass squared differences of active neutrinos, two sterile neutrino masses and active-sterile neutrino mixing angle versus $U(1)_{B-L}$ gaugino mass $m_{BL}$, where (a) the solid line stands for $\sin^2{\theta_{23}}$ versus $m_{BL}$, the dashed line stands for $\sin^2{\theta_{12}}$ versus $m_{BL}$, and the dotted line stands for $\sin^2{\theta_{13}}$ versus $m_{BL}$, (b) the solid line stands for $\Delta{m_A^2}$ versus $m_{BL}$ and the dashed line stands for $\Delta{m_{\odot}^2}$ versus $m_{BL}$, (c) the solid line stands for $m_{r1}$ versus $m_{BL}$ and the dashed line stands for $m_{r2}$ versus $m_{BL}$, and (d) the solid line stands for $\sin^2\theta_1$ versus $m_{BL}$.}
\label{fig5}
\end{figure}
 Taking $\tan\beta=20$, $\upsilon_{N_1}=3\;\rm{GeV}$, and $\upsilon_{N_2}=7.7\;\rm{GeV}$, we plot the active neutrino mixing angles varying with $m_{BL}$ in Fig. 5(a), where the solid line denotes  $\sin^2{\theta_{23}}$ versus $\upsilon_{N_2}$, the dashed line denotes $\sin^2{\theta_{12}}$ versus $\upsilon_{N_2}$, and the dotted line denotes $\sin^2{\theta_{13}}$ versus $\upsilon_{N_2}$. With the increasing of $m_{BL}$, the theoretical predictions on the mixing angles of active neutrinos $\sin^2{\theta_{23}}$ and $\sin^2{\theta_{13}}$ depend on $m_{BL}$ mildly, and the mixing angle $\sin^2{\theta_{12}}$ decreases steeply. Adopting the same choice on parameter space, we plot the mass squared differences of active neutrinos varying with $m_{BL}$ in Fig. 5(b), where the solid line denotes $\Delta{m_A^2}$ versus $m_{BL}$ and the dashed line denotes $\Delta{m_{\odot}^2}$ versus $m_{BL}$. It shows that the mass squared differences of active neutrinos $\Delta{m_A^2}$ and $\Delta{m_{\odot}^2}$ increase steeply with the increasing of $m_{BL}$. From those numerical results, we find that the updated experimental data require $m_{BL}\sim1.08\;\rm{TeV}$. In addition, we study the masses of two sterile neutrinos varying with  $\upsilon_{N_2}$ in Fig. 5(c), where the solid line denotes $m_{r1}$ versus $m_{BL}$ and the dashed line denotes $m_{r2}$ versus $m_{BL}$. It shows that the two sterile neutrino masses have the same trend of variability with the NO spectrum. Under the same choice on parameter space, the numerical result of the active-heavier sterile neutrino mixing angle $\sin^2\theta_2$ changes gently about $10^{-10}$. We only study the active-sterile neutrino mixing angle $\sin^2\theta_1$ versus $m_{BL}$ in Fig. 5(d). With increasing of $m_{BL}$, the active-sterile neutrino mixing angle $\sin^2\theta_1$ increases gently.
\begin{figure}[h]
\setlength{\unitlength}{1mm}
\centering
\includegraphics[width=2.5in]{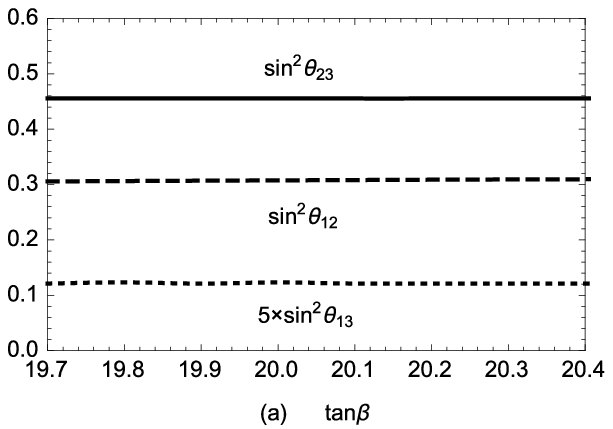}
\vspace{0.0cm}
\includegraphics[width=2.5in]{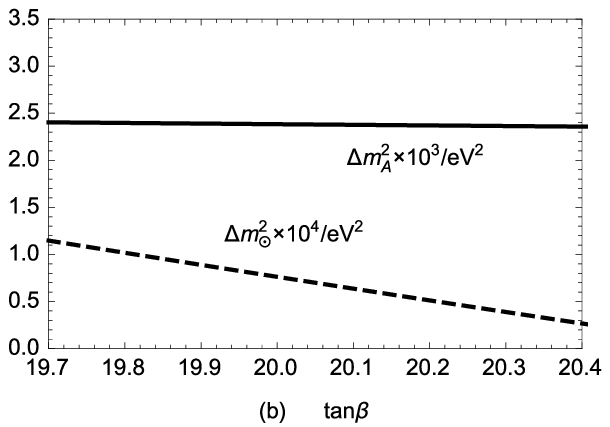}
\vspace{0cm}
\includegraphics[width=2.5in]{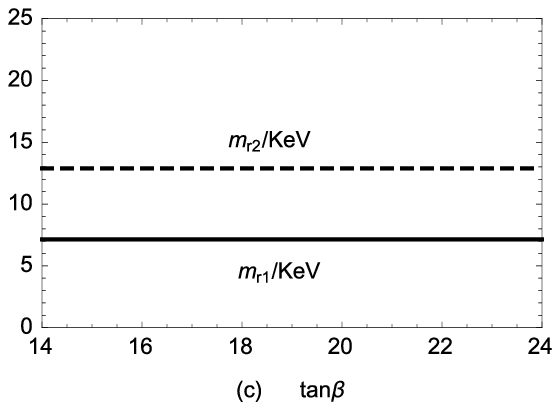}
\vspace{0cm}
\includegraphics[width=2.5in]{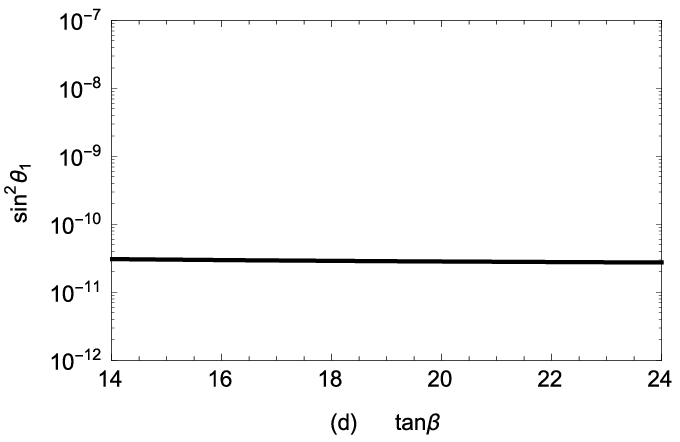}
\vspace{0cm}
\caption[]{Assuming neutrino mass spectrum with IO, we plot the mixing angles, mass squared differences of active neutrinos, two sterile neutrino masses and active-sterile neutrino mixing angle versus $\tan\beta$, where (a) the solid line stands for $\sin^2{\theta_{23}}$ versus $\tan\beta$, the dashed line stands for $\sin^2{\theta_{12}}$ versus $\tan\beta$, and the dotted line stands for $\sin^2{\theta_{13}}$ versus $\tan\beta$, (b) the solid line stands for $\Delta{m_A^2}$ versus $\tan\beta$ and the dashed line stands for $\Delta{m_{\odot}^2}$ versus $\tan\beta$, (c) the solid line stands for $m_{r1}$ versus $\tan\beta$ and the dashed line stands for $m_{r2}$ versus $\tan\beta$, and (d) the solid line stands for $\sin^2\theta_1$ versus $\tan\beta$.}
\label{fig6}
\end{figure}

Taking $m_{BL}=1.08\;\rm{TeV}$, $\upsilon_{N_1}=3\;\rm{GeV}$, and $\upsilon_{N_2}=7.7\;\rm{GeV}$, we draw the neutrino mixing angles varying with  $\tan\beta$ in Fig. 6(a). With the increasing of $\tan\beta$, theoretical predictions on those mixing angles vary gently. Adopting the same choice on parameter space, we draw the mass squared differences of active neutrinos varying with $\tan\beta$ in Fig. 6(b). It shows that $\Delta{m_A^2}$ (solid line) changes gently with the increasing of $\tan\beta$; however, $\Delta{m_{\odot}^2}$ (dashed line) decreases rapidly with the increasing of $\tan\beta$. In addition, we study the masses of two sterile neutrinos versus $\tan\beta$ in Fig. 6(c). It shows that the masses of two sterile neutrinos depend on $\tan\beta$ gently. The active-sterile neutrino mixing angles have barely changed with $\tan\beta$. The numerical result of $\sin^2\theta_2$ is about $10^{-10}$. The numerical result of $\sin^2\theta_1$ is about $10^{-11}$ in Fig. 6(d).

Assuming the neutrino mass spectrum with inverted ordering, two sterile neutrinos obtain $\rm{KeV}$ scale masses. The lighter sterile neutrino forms dark matter, and the heavier sterile neutrino can decay into the lighter one to enrich the production of the sterile neutrino DM.

\section{summary\label{sec6}}

We investigate the origin of neutrino masses in the MSSM with local $U(1)_{B-L}$ symmetry. In this model sneutrinos all obtain nonzero VEVs. We constrain the relevant parameter space by the neutrino oscillation experimental data and the mass of the lightest \textit{CP}-even Higgs. At tree level, three left-handed neutrinos and two sterile neutrinos obtain masses through the seesaw mechanism, but the masses of two sterile neutrinos are very tiny. The one-loop radiation corrections to the mass matrix of neutralino-neutrino are also studied.  Both NO spectrum and IO spectrum are studied. The active neutrino mass squared differences and mixing angles can account for the experimental data on neutrino oscillations. Because the one-loop radiative corrections to the left-handed neutrinos are suppressed by the tiny Yukawa couplings and the small nonzero VEVs of left-handed sneutrinos, the corrections to the active neutrinos are very small. The active neutrinos obtain mass mainly from tree level. When one-loop corrections are included, the numerical results show that there is parameter space to give two sterile neutrinos with $\rm{KeV}$ masses and the small active-sterile neutrino mixing angle. When $\upsilon_{N_2}=7.7\;\rm{GeV}$, $m_{BL}=1.08\;\rm{TeV}$, and $\tan\beta=20$, the mass of the heavier sterile neutrino $m_{r2}$ is about $12.88\;\rm{KeV}$, the mixing angle $\sin^2\theta_2$ is about $10^{-10}$, the mass of lighter sterile neutrino $m_{r1}$ is about $7.1\;{\rm{KeV}}$, and the mixing angle $\sin^2\theta_1$ is about $10^{-11}$. The lighter sterile neutrino can account for the observed X-ray line at $3.5\;{\rm{KeV}}$ \cite{sterile neutrino1,sterile neutrino2}. Therefore, the lighter sterile neutrino can be a dark matter candidate. However the oscillation mechanism does not produce enough of these neutrinos to act as all dark matter \cite{neutrino DM,DW}. Nonresonant production contributes to the dark matter abundance $\Omega_{s}h^2\approx0.3({{\sin^2{2\theta}}\over{10^{-10}}})({{m_r}\over{100{\rm{kev}}}})^2$  \cite{neutrino DM}. Only $1\%$ of dark matter is produced for the lightest sterile neutrino. In this model, the heavier sterile neutrino can decay into the lighter one to enrich the production of the sterile neutrino DM. However, this still does not produce enough dark matter and it may require other mechanisms (Shi-Fuller mechanism \cite{ShF}) in the early Universe \cite{DM}.

\begin{acknowledgments}
\indent\indent
The work has been supported by the National Natural Science Foundation of China (NNSFC)
with Grants No. 11275036, No. 11535002, No. 11647120, and No. 11705045,
the Natural Science Foundation of Hebei province with Grants No.A2016201010, and A2016201069,
the Natural Science Foundation of Hebei University with Grants No. 2011JQ05, and No. 2012-242,
and the Hebei Key Lab of Optic-Electronic Information and Materials.
\end{acknowledgments}

\appendix

\section{The effective mass matrices for right-handed sneutrinos\label{app1}}
\indent\indent
Considering the last minimization condition in Eq. (8), the most general matrix is
\begin{eqnarray}
&&m_{\tilde{N}^c}^2=\left(\begin{array}{ccc}
\xi_{\tilde{N}_1^c}^2-{m_{Z_{BL}}^2/2} & {-\xi_{\tilde{N}_1^c}^2{\upsilon_{N_1}}-\xi_{\tilde{N}_2^c}^2{\upsilon_{N_2}}+\xi_{\tilde{N}_3^c}^2{\upsilon_{N_3}}}\over{2{\upsilon_{N_1}}{\upsilon_{N_2}}} & {-\xi_{\tilde{N}_1^c}^2{\upsilon_{N_1}}+\xi_{\tilde{N}_2^c}^2{\upsilon_{N_2}}-\xi_{\tilde{N}_3^c}^2{\upsilon_{N_3}}}\over{2{\upsilon_{N_1}}{\upsilon_{N_3}}}\\
{-\xi_{\tilde{N}_1^c}^2{\upsilon_{N_1}}-\xi_{\tilde{N}_2^c}^2{\upsilon_{N_2}}+\xi_{\tilde{N}_3^c}^2{\upsilon_{N_3}}}\over{2{\upsilon_{N_1}}{\upsilon_{N_2}}} & \xi_{\tilde{N}_2^c}^2-{m_{Z_{BL}}^2/2} & {\xi_{\tilde{N}_1^c}^2{\upsilon_{N_1}}-\xi_{\tilde{N}_2^c}^2{\upsilon_{N_2}}-\xi_{\tilde{N}_3^c}^2{\upsilon_{N_3}}}\over{2{\upsilon_{N_2}}{\upsilon_{N_3}}}\\
{-\xi_{\tilde{N}_1^c}^2{\upsilon_{N_1}}+\xi_{\tilde{N}_2^c}^2{\upsilon_{N_2}}-\xi_{\tilde{N}_3^c}^2{\upsilon_{N_3}}}\over{2{\upsilon_{N_1}}{\upsilon_{N_3}}} &  {\xi_{\tilde{N}_1^c}^2{\upsilon_{N_1}}-\xi_{\tilde{N}_2^c}^2{\upsilon_{N_2}}-\xi_{\tilde{N}_3^c}^2{\upsilon_{N_3}}}\over{2{\upsilon_{N_2}}{\upsilon_{N_3}}} & \xi_{\tilde{N}_3^c}^2-{m_{Z_{BL}}^2/2}\end{array}\right),
\label{mN2}
\end{eqnarray}
with $\xi_{\tilde{N}_1^c}^2=(m_{\tilde{N}^c}^2)_{11}+{m_{Z_{BL}}^2/2}$, $\xi_{\tilde{N}_2^c}^2=(m_{\tilde{N}^c}^2)_{22}+{m_{Z_{BL}}^2/2}$, and $\xi_{\tilde{N}_3^c}^2=(m_{\tilde{N}^c}^2)_{33}+{m_{Z_{BL}}^2/2}$. There are three parameters in matrix $m_{\tilde{N}^c}^2$. We make an approximation of this matrix to reduce the number of free parameters. A possible solution is that $\upsilon_{N_1}$ and $\upsilon_{N_2}$ are small and $\upsilon_{N_3}$ is large for large $\upsilon_N$ ($\upsilon_N^2=\sum\limits_{\alpha=1}^3\upsilon_{N_\alpha}^2$). In this case, we can obtain $(m_{\tilde{N}^c}^2)_{33}\simeq-{m_{Z_{BL}}^2/2}$ from the last minimization condition in Eq. (8). Compared to  $(m_{\tilde{N}^c}^2)_{33}=\xi_{\tilde{N}_3^c}^2-{m_{Z_{BL}}^2/2}$ in Eq. (B1), $\xi_{\tilde{N}_3^c}^2$ should be small. If we select the appropriate parameters $\xi_{\tilde{N}_{1,2}^c}^2$ and choose  $\xi_{\tilde{N}_3^c}^2={{\xi_{\tilde{N}_1^c}^2{\upsilon_{N_1}}+\xi_{\tilde{N}_2^c}^2{\upsilon_{N_2}}}\over{\upsilon_{N_3}}}$, $\xi_{\tilde{N}_3^c}^2$ should be small considering small $\upsilon_{N_{1,2}}$ and large $\upsilon_{N_3}$. The number of matrix parameters is reduced to two. Then, we can have Eq.(9).

Using the minimization conditions, we derive the $3\times3$ mass squared matrix for neutral \textit{CP}-odd scalars $P_{\tilde{N}_I}^0$ at tree level
\begin{eqnarray}
&&({M_{\tilde{N}^c}^2}^{odd})_{IJ}=(m_{\tilde{N}^c}^2)_{IJ}+{{m_{Z_{BL}}^2}\over2}\delta_{IJ},
\label{CP-odd}
\end{eqnarray}
with $I,J=1,2,3$ denoting the index of generation. Correspondingly the orthogonal $3\times3$ matrix from interaction eigenstates to mass eigenstates is written as $U_{\tilde{N}_O}$, and three masses of the  \textit{CP}-odd scalars are
\begin{eqnarray}
&&m_{P_{\tilde{N}_1}}^2=0,\qquad m_{P_{\tilde{N}_2}}^2={{\omega_A-\omega_B}\over{2\upsilon_{N_3}^2}},\qquad m_{P_{\tilde{N}_3}}^2={{\omega_A+\omega_B}\over{2\upsilon_{N_3}^2}}.
\label{CP-odd-orthogonal}
\end{eqnarray}
and the concrete expressions for $\omega_{A,B}$ are
\begin{eqnarray}
&&\omega_A=\xi_{\tilde{N}_1^c}^2(\upsilon_N^2-\upsilon_{N_2}^2)+
\xi_{\tilde{N}_2^c}^2(\upsilon_N^2-\upsilon_{N_1}^2),\nonumber\\
&&\omega_B^2=\omega_A^2-4\xi_{\tilde{N}_1^c}^2\xi_{\tilde{N}_2^c}^2\upsilon_N^2\upsilon_{N_3}^2.
\label{CP-odd-orthogonal}
\end{eqnarray}
Additionally the $3\times3$  mass squared matrix for neutral \textit{CP}-even scalars $\tilde{\nu}_{R_I}$ is
\begin{eqnarray}
&&({M_{\tilde{N}^c}^2}^{even})_{IJ}=(m_{\tilde{N}^c}^2)_{IJ}+{{m_{Z_{BL}}^2}\over2}\delta_{IJ}+g_{BL}^2\upsilon_{N_I}\upsilon_{N_J}.
\label{CP-even}
\end{eqnarray}
Correspondingly the orthogonal $3\times3$ matrix is written as $U_{\tilde{N}_E}$, and three masses of the  \textit{CP}-even scalars are
\begin{eqnarray}
&&m_{H_{\tilde{N}_1}}^2=m_{Z_{BL}}^2,\qquad m_{H_{\tilde{N}_2}}^2={{\omega_A-\omega_B}\over{2\upsilon_{N_3}^2}},\qquad m_{H_{\tilde{N}_3}}^2={{\omega_A+\omega_B}\over{2\upsilon_{N_3}^2}}.
\label{CP-even-orthogonal}
\end{eqnarray}

\section{The effective mass matrix for five light neutrinos at tree level\label{app2}}
\indent\indent
The effective mass matrix for five light neutrinos at tree level is
\begin{eqnarray}
&&m_\nu^{eff}\simeq\left(\begin{array}{cc}
[M_\nu^{LL}]_{3\times3} & [M_\nu^{LR}]_{3\times2}\\
{[M_\nu^{LR,T}]_{2\times3}} & [M_\nu^{RR}]_{2\times2}
\end{array}\right),
\label{meff}
\end{eqnarray}
where
\begin{eqnarray}
&&{(M_\nu^{LL})_{ij}}={\upsilon_{L_i}\upsilon_{L_j}\over{\Lambda_\upsilon}}+
{{\upsilon_{L_i}\zeta_j+\upsilon_{L_j}\zeta_i}\over{\Lambda_{\upsilon\zeta}}}+{{\zeta_i\zeta_j}\over{\Lambda_\zeta}},\nonumber\\
&&(M_\nu^{LR})_{i1}=\delta_{i3},\qquad\qquad\qquad\quad  (M_\nu^{LR})_{i2}=\delta_{i2},\nonumber\\
&&{(M_\nu^{RR})_{11}}={{\tilde{m}(m_{_{BL}}^2-\Delta_{_{BL}}^2)\upsilon_d^2}\over{\Lambda_{\tilde{m}^4}}}\varepsilon_{13}^2,\nonumber\\
&&{(M_\nu^{RR})_{12}}={{\tilde{m}(m_{_{BL}}^2-\Delta_{_{BL}}^2)\upsilon_d^2}\over{\Lambda_{\tilde{m}^4}}}\varepsilon_{12}\varepsilon_{13},\nonumber\\
&&{(M_\nu^{RR})_{22}}={{\tilde{m}(m_{_{BL}}^2-\Delta_{_{BL}}^2)\upsilon_d^2}\over{\Lambda_{\tilde{m}^4}}}\varepsilon_{12}^2,
\label{meff}
\end{eqnarray}
with
\begin{eqnarray}
&&{1\over{\Lambda_\upsilon}}={{(m_{_{BL}}^2-\Delta_{_{BL}}^2)\tilde{m}}\over{\Lambda_{\tilde{m}^4}}}\mu^2,\nonumber\\
&&{1\over{\Lambda_\zeta}}={{2\tilde{\mu}^4\upsilon_u^2{m_{_{BL}}}}\over{\Lambda_{\tilde{m}^4}\upsilon_N^2}}+
{{(m_{_{BL}}^2-\Delta_{_{BL}}^2)\tilde{m}}\over{2\Lambda_{\tilde{m}^4}}}\upsilon_d^2,\nonumber\\
&&{1\over{\Lambda_{\upsilon\zeta}}}={{{\sqrt2}\tilde{\mu}^4{g_{BL}^2}\upsilon_u}\over{\Lambda_{\tilde{m}^4}}}+
{{(m_{_{BL}}^2-\Delta_{_{BL}}^2)\mu\tilde{m}}\over{{\sqrt2}\Lambda_{\tilde{m}^4}}}\upsilon_d,\nonumber\\
&&\Lambda_{\tilde{m}^4}=2(\Delta_{_{BL}}^2-m_{_{BL}}^2)\tilde{\mu}^2+(\Delta_{_{BL}}+m_{_{BL}})\tilde{m}\upsilon_d^2\varepsilon_-^2
+(m_{_{BL}}-\Delta_{_{BL}})\tilde{m}\upsilon_d^2\varepsilon_+^2,\nonumber\\
&&\tilde{m}={1\over2}(g_1^2{M_2}+g_2^2{M_1}),\nonumber\\
&&{\tilde{\mu}^4}=M_1M_2\mu^2-\tilde{m}\mu\upsilon_d\upsilon_u.
\label{abbreviation}
\end{eqnarray}

\section{Radiative corrections on masses of sterile neutrinos\label{app3}}
\indent\indent
As $\alpha=1,2$ and $\beta=3,4$, the renormalized self-energy is formulated as
\begin{eqnarray}
&&\hat{\Sigma}_{\alpha\beta}^L(p^2)=\hat{\Sigma}_{\alpha\beta}^{L(1)}(p^2)+\hat{\Sigma}_{\alpha\beta}^{L(2)}(p^2),\nonumber\\
&&\hat{\Sigma}_{\alpha\beta}^M(p^2)=\hat{\Sigma}_{\alpha\beta}^{M(1)}(p^2)+\hat{\Sigma}_{\alpha\beta}^{M(2)}(p^2).
\label{M1234}
\end{eqnarray}
In view of $(U_N)_{41}=(U_N)_{42}=0$, and $m_{N_{1,2}}\ll{m_{N_{3,4}}}$, the corrections to the mass matrix are
\begin{eqnarray}
&&(\Delta{\cal M}_N^{(0)})_{\alpha\beta}={g_{BL}^2\over{(4\pi)^2}}\sum_{\delta=3}^4\sum_{i,j}^3m_{N_\delta}
\{{\cal{R}}((U_N)_{i\alpha}(U_N)_{4\delta}(U_N)_{j\delta}(U_N)_{4\beta})\nonumber\\
&&\hspace{2.2cm}\times[\sum_{a=1}^3(U_{\tilde{N}_E})_{ia}(U_{\tilde{N}_E})_{ja}\hat{B}_0(m_{N_\alpha}^2,m_{N_\delta}^2,m_{H_{N_a}}^2)\nonumber\\
&&\hspace{2.2cm}-\sum_{a=2}^3(U_{\tilde{N}_O})_{ia}(U_{\tilde{N}_O})_{ja}\hat{B}_0(m_{N_\alpha}^2,m_{N_\delta}^2,m_{P_{N_a}}^2)]\nonumber\\
&&\hspace{2.2cm}+{\cal{R}}(2(U_N)_{i\alpha}(U_N)_{i\delta}^\ast(U_N)_{j\delta}^\ast(U_N)_{j\beta})
(\hat{B}_0-{1\over2})(m_{N_\alpha}^2,m_{N_\delta}^2,m_{Z_{BL}}^2)\nonumber\\
&&\hspace{2.2cm}+{\cal{R}}(2(U_N)_{i\beta}(U_N)_{i\delta}^\ast(U_N)_{j\delta}^\ast(U_N)_{j\alpha})
(\hat{B}_0-{1\over2})(m_{N_\beta}^2,m_{N_\delta}^2,m_{Z_{BL}}^2)\}\nonumber\\
&&\hspace{2.2cm}-{g_{BL}^2\over{(4\pi)^2}}m_{N_\beta}\sum_{\delta=1}^4\sum_{i,j}^3{\cal{R}}
((U_N)_{i\beta}^\ast(U_N)_{i\delta}^\ast(U_N)_{j\delta}^\ast(U_N)_{j\alpha})\nonumber\\
&&\hspace{3.0cm}\times(\hat{B}_1-{1\over2})(m_{N_\beta}^2,m_{N_\delta}^2,m_{Z_{BL}}^2).
\label{MN1234}
\end{eqnarray}
As $\alpha,\beta=3,4$, the result of one-loop corrections to the mass matrix is
\begin{eqnarray}
&&(\Delta{\cal M}_N^{(0)})_{\alpha\beta}={g_{BL}^2\over{(4\pi)^2}}\sum_{\delta=3}^4\sum_{i,j}^3m_{N_\delta}
\{{\cal{R}}((U_N)_{i\alpha}(U_N)_{4\delta}(U_N)_{j\delta}(U_N)_{4\beta})\nonumber\\
&&\hspace{2.2cm}\times[\sum_{a=1}^3(U_{\tilde{N}_E})_{ia}(U_{\tilde{N}_E})_{ja}\hat{B}_0(m_{N_\alpha}^2,m_{N_\delta}^2,m_{H_{N_a}}^2)\nonumber\\
&&\hspace{2.2cm}-\sum_{a=2}^3(U_{\tilde{N}_O})_{ia}(U_{\tilde{N}_O})_{ja}\hat{B}_0(m_{N_\alpha}^2,m_{N_\delta}^2,m_{P_{N_a}}^2)]\nonumber\\
&&\hspace{2.2cm}+{\cal{R}}(2(U_N)_{i\alpha}(U_N)_{i\delta}^\ast(U_N)_{j\delta}^\ast(U_N)_{j\beta})
(\hat{B}_0-{1\over2})(m_{N_\alpha}^2,m_{N_\delta}^2,m_{Z_{BL}}^2)
+(\alpha\leftrightarrow\beta)\}\nonumber\\
&&\hspace{2.2cm}-{g_{BL}^2\over{(4\pi)^2}}\sum_{\delta=1}^4\sum_{i,j}^3\{{m_{N_\alpha}}
{\cal{R}}((U_N)_{i\alpha}^\ast(U_N)_{4\delta}^\ast(U_N)_{j\delta}(U_N)_{4\beta})\nonumber\\
&&\hspace{2.2cm}\times[\sum_{a=1}^3(U_{\tilde{N}_E})_{ia}(U_{\tilde{N}_E})_{ja}\hat{B}_1(m_{N_\alpha}^2,m_{N_\delta}^2,m_{H_{N_a}}^2)\nonumber\\
&&\hspace{2.2cm}+\sum_{a=1}^3(U_{\tilde{N}_O})_{ia}(U_{\tilde{N}_O})_{ja}\hat{B}_1(m_{N_\alpha}^2,m_{N_\delta}^2,m_{P_{N_a}}^2)]\nonumber\\
&&\hspace{2.2cm}+{\cal{R}}((U_N)_{i\alpha}^\ast(U_N)_{i\delta}^\ast(U_N)_{j\delta}^\ast(U_N)_{j\beta})(\hat{B}_1-{1\over2})(m_{N_\alpha}^2,m_{N_\delta}^2,m_{Z_{BL}}^2)\nonumber\\
&&\hspace{2.2cm}+(\alpha\leftrightarrow\beta)\}.
\label{MN34}
\end{eqnarray}
The corrections from real and image components of right-handed sneutrinos only appear as $\beta=3,4$, so their contributions to the sterile neutrino masses are relatively small.

\end{document}